\documentclass[11pt]{article}

\usepackage{amstext,amsmath,amssymb,amsfonts,bbm}
\usepackage{hyperref}
\usepackage{authblk}
\usepackage{caption}
\usepackage{subcaption}
\usepackage[utf8]{inputenc}
\usepackage{epsfig}
\usepackage{amsthm}
\usepackage{mathtools}
\usepackage{xcolor}
\usepackage{graphicx}
\usepackage{braket}
\usepackage{slashed}
\usepackage{float}
\usepackage{enumitem}
\usepackage{physics}

\usepackage[lmargin=80pt,rmargin=80pt,tmargin=80pt,bmargin=80pt]{geometry}

\usepackage{comment}
\usepackage{esint}
\usepackage[normalem]{ulem}

\newcommand{\be}{\begin{equation}}
\newcommand{\ee}{\end{equation}}

\newcommand{\f}{\frac}
\newcommand{\p}{\partial}

\theoremstyle{remark}

\let\a=\alpha \let\b=\beta  \let\g=\gamma  \let\d=\delta
        \let\l=\lambda
\let\m=\mu    \let\n=\nu           
 \let\t=\tau     
    
     \let\X=F

\newcommand{\sgn}{\text{sgn}}

\newcommand{\beq}{\begin{equation}}
\newcommand{\eeq}{\end{equation}}
\newcommand{\bea}{\begin{eqnarray}}
\newcommand{\eea}{\end{eqnarray}}

\newcommand{\al}[1]{\begin{align}#1\end{align}}
\newcommand{\sD}{\slashed{D}}
\newcommand{\DG}{\Delta_{\Gamma}}
\newcommand{\nnn}{ \nonumber \\}
\newcommand{\iif}{\,\quad \text{if}\,\,}

\begin{document}

\title{\bf Dirac walks on regular trees}

\author[1]{Nicolas Delporte}
\author[1]{Saswato Sen}
\author[1]{Reiko Toriumi}

\affil[1]{\normalsize\it Okinawa Institute of Science and Technology Graduate University, 1919-1, Tancha, Onna, Kunigami District, Okinawa 904-0495, Japan. \authorcr
Emails: {\rm\url{nicolas.delporte@oist.jp}, \url{saswato.sen@oist.jp}, \url{reiko.toriumi@oist.jp} }\authorcr \hfill }


\date{\vspace{-5ex}}

\maketitle

\hrule\bigskip

\begin{abstract}

\noindent 
The study of matter fields on an ensemble of random geometries is a difficult problem still in need of new methods and ideas. 
We will follow a point of view inspired by probability theory techniques that relies on an expansion of the two point function as a sum over random walks.
An analogous expansion for Fermions on non-Euclidean geometries is still lacking. Casiday et al. [\textit{Laplace and Dirac operators on
graphs}, Linear and Multilinear Algebra (2022) 1] proposed a classical “Dirac walk” diffusing on vertices and edges of an oriented graph with a square root of the graph Laplacian. In contrast to the simple random walk, each step of the walk is given a sign depending on the orientation of the edge it goes through. In a toy model, we propose here to study the Green functions, spectrum and the spectral dimension of such “Dirac walks” on the Bethe lattice, a $d$-regular tree. The recursive structure of the graph makes the problem exactly solvable. Notably, we find that the spectrum develops a gap and that the spectral dimension of the Dirac walk matches that of the simple random walk ($d_s=1$ for $d=2$ and $d_s=3$ for $d\geq 3$).

\end{abstract}

\hrule\bigskip

\tableofcontents

\section{Introduction}
\label{sec:intro}

Our work attempts to tie stronger bonds between combinatorics, probabilistic methods and quantum field theories (QFTs), in order to understand the coupling between matter and geometry when the latter is a general manifold, obtained as a continuous limit of graphs. 
Such discrete structures could be randomly generated, with probability distributions arising from gravity \footnote{Discrete models of gravity have been explored in the context of causal and Euclidean dynamical triangulations, group field theory, Regge calculus, spin foams, matrix models, tensor models, etc. \cite{Loll_2019, ambjorn2022elementary,deboer2022frontiers,Williams:2006kp}},
or to account for some disorder in the system, leading for example to localization phenomena \cite{anderson1958absence}.
Two dimensions have observed most of the progress in this program of probabilistic interpretation of quantum gravity, through various matrix models whose Feynman diagrams represent discretized surfaces, furnishing conjectures for how critical matter behaves in the presence of a fluctuating $2d$ geometry \cite{David:1985nj, Knizhnik:1988ak, DISTLER1989509}, culminating later with a probabilistic proof of the KPZ exponents \cite{duplantier2011liouville} through the Gaussian free field that has sparked a vast area of research (see \cite{ding2021introduction} for an overview). 

In quantum mechanics, after Feynman and Kac expressed path integrals with a Brownian motion, Symanzik \cite{Symanzik:1968zz} proposed to use this stochastic reformulation of the path integral for the purposes of constructive QFT \cite{jaffe2000constructive}.
In particular, correlation functions translate into intersection probabilities of classical random walks. Simultaneously, random walks have been used as a perturbative yet a powerful tool to prove a large number of results in equilibrium statistical physics: the triviality of $\phi^4$ in $d\geq 4$ \cite{aizenman2021marginal}, bounds on the critical exponents of lattice spin systems \cite{fernandez2013random}, etc. 
In that respect, different processes related to Markov chains have been introduced to accommodate the different models (self-avoiding walks, loop-erased random walks, etc.) \cite{lawler12}.

We would like to develop similar tools to study quantum fields with spin, keeping in mind applications to general geometries (fractals, random or irregular lattices, etc). 
Various extensions of the path integral formulation to the spinorial case have already been suggested (see \cite{Bodmann_1999} and references therein). 
Essentially, they amount to including a functional that couples the spin to the tangent vector of the path of the random walk.
A notable difference from the spinless case is that a spin term in the Hamiltonian changes the Hausdorff dimension of the continuous paths which are summed over, from $d_H=2$ to $d_H=1$, hence smooths the path.
Analogous discussions \cite{jaroszewicz1994random} have shown that this difference explains also the reduced critical dimension of quartic Fermions ($d_c=2$) compared to that of Bosons ($d_c=4$).

Another notion of dimension that plays a relevant role in fractal and random graphs is the spectral dimension. 
It was introduced by Alexander and Orbach \cite{alexander1982density} to describe the low frequency vibrational spectrum and it relates to the long time behavior of the return probability of a random walker. 
A recently proven conjecture states that the spectral dimension of a random walk on percolation clusters on $\mathbb{Z}^d$ with large $d$ is $4/3$ (the result does not hold for $d<6$)  \cite{kozma2009alexander}. 
This result stands also valid for the Infinite Incipient Cluster on trees \cite{barlow2006random}, giving a sense for which an infinite tree is a large dimensional background, displaying mean-field properties. 
The determination of the spectral dimension on diverse graphs necessitates a good control of the propagating kernel of the random walker. 
Cassi et al. \cite{burioni1996universal} proved several universal properties of the spectral dimension, such as its independence of an additional waiting time of the walker and of a fluctuating mass on the lattice. 
Delporte et al. \cite{delporte2021perturbative} showed that it is central in the study of the renormalization group of a long-range scalar field theory. 

In order to discuss Fermi-Dirac statistics, the simplest starting point is spin $1/2$, with anti-commuting fields obeying the Dirac equation in flat space. Setting the problem on a discrete graph boils down to finding a Fermion discretization with the correct continuous limit and number of degrees of freedom.
Dirac Fermions have been discretized in a variety of ways, most of which suffer from the Fermion doubling problem \cite{montvay1994quantum}. One combinatorial approach to Grassmann fields defined on vertices of general graphs leads to exact expressions of partition and correlation functions in terms of sums over forests over the graph \cite{Caracciolo:2004hz}.
The K\"ahler-Dirac approach particularly suits simplicial graphs \cite{knill2013dirac} \footnote{See \cite{catterall2018kahler} for their implementation on Euclidean dynamical triangulations.} \footnote{On a hypercubic lattice and after relabeling, this formalism reduces to Susskind's staggered Fermions. 
Notice that the K\"ahler-Dirac equation behaves differently from the covariant Dirac equation when curvature is involved \cite{banks1982geometric}. }
and their simplest version, trees, that will be our main concern. 

Casiday et al. \cite{casiday2022laplace} proposed an expansion of powers of a formal Dirac operator (i.e. square root of the graph Laplacian), whose quadratic form matches the K\"ahler-Dirac action on trees, in terms of superwalks and vertex-edge walks. Below both types are called Dirac walks, that is, random walks on vertices and edges of a directed graph taking up a sign depending on the orientation of the edge it traverses.

To simplify our discussion, we will only focus on tree graphs of regular degree $d$, hence without boundary, known as the Bethe lattice, on which many statistical models are exactly solvable \cite{baxter1982exactly}. 
They serve as an effectively large dimensional testbed to understand non-trivial physics, such as Anderson localization \cite{pascazio2023anderson}, glasses \cite{rivoire2004glass}, etc.
The Bethe lattice corresponds also to the large $f$ limit of $(f,d)$ tesselations \footnote{Here, $f$ corresponds to the number of sides of the polygons and $d$ to the number of polygons joining around every vertex.} of the hyperbolic plane \cite{mosseri1982bethe}. 
Statistical models defined on those lattices with exponential growth of number of sites present interesting critical behaviors, with multiple phase transitions and mean-field critical exponents \cite{Breuckmann_2020}, but also an uncountable set of Gibbs states \cite{wu2000ising,gandolfo2015manifold}.
Inspiring sources for the simple random walk on the Bethe lattice were the references \cite{hughes1982random} and \cite{monthus1996random}. In the first, they used the generating function formalism to derive statistics of the process, mapping the Bethe lattice to the half-line. In \cite{monthus1996random}, in addition to a comparison with a random walk in hyperbolic geometry, they study the asymptotics of the walk, relevant to retrieve the spectral dimension.

Let us emphasize that, in our study, the Dirac walk is an entirely classical process; however since the weights associated to the walks can be positive and negative, it cannot be given a probabilistic interpretation.

At the end of our work, we learned that a similar construction of the Dirac operator on general graphs (networks) was written in \cite{Bianconi_2021,Bianconi_2023,Bianconi_2024}, generalizing the graph Dirac operator to higher simplices and \textit{multiplex networks}. Their formalism allowed to study extensively the energy spectrum of topological Dirac Fermions and its interplay with the geometry of the network (see \cite{nokkala2023complex} for a recent overview of their results and research). Our approach gives rather a combinatorial and analytic expression of the spectrum of such theories on the Bethe lattice.\newline

We plan our discussions as follows:
We will start in Section~\ref{ch-1} by describing a framework to study quantum fields on general graphs. 
The propagation will be given by the square root of the Laplacian, in the formulation proposed by \cite{casiday2022laplace}. 
In the next Section~\ref{ch-2}, specializing to the Bethe lattice, we will compute the generating function of weighted walks from vertices or edges. From their asymptotics, we will derive the associated spectral dimension of the walk, reminding in parallel the spectral dimension of a simple random walk. 
Section \ref{ch-5} will be devoted to Green's functions and spectral densities on the Bethe lattice with $d\geq 2$, obtained directly from the preceding generating functions. We will compare again our results to those of the simple walks and discuss the observed analytic properties.
In Section~\ref{sec:concl}, we will present a brief summary and pose questions to explore in the future.

\section{Quantum field theory on graphs}
\label{ch-1}

In this section, we will review how to formulate quantum field theories (QFT) on graphs in the Euclidean path integral formalism. 
In Section \ref{sec:SFTonGraphs}, we will initially review a free scalar QFT on a graph and comment on the relation between the two-point function and random walk transition probabilities on graphs.
Then, in Section~\ref{subsec:FFeuclid}, we will review necessary aspects of Fermions in $\mathbb{R}^n$ to describe a Fermionic action on graphs and discuss the difficulties of defining a Dirac operator on general graphs. We will attempt to resolve this problem by defining a particular Dirac operator that has a combinatorial interpretation in terms of a particular kind of signed walk, which we will call the `Dirac walk' in the Section~\ref{sec:FermionicQFt}, as suggested by \cite{casiday2022laplace}. 
In the same Section~\ref{sec:FermionicQFt}, we will illustrate the connection between the Dirac walk and the two-point function of a free Fermionic action.

\subsection{Free scalars on graphs}
\label{sec:SFTonGraphs}

We define a digraph (directed graph) $\Gamma(V,E)$ with vertices $V(\Gamma)$ and directed edges $E(\Gamma)=\{e = (v,w)~\rvert~ v,w \in V(\Gamma)\}$, where $(v,w)$ is an ordered pair of vertices, with orientation going from $v$ to $w$. 
The cardinality of $V(\Gamma)$ will be written as $|V|$,  and of $E(\Gamma)$ will be represented by $|E|$. \footnote{$|V|$ and $|E|$ can be infinite.}
The incidence matrix of the digraph is a $|V|\times |E|$ matrix defined such that 
\begin{align}
\label{eq:Incident}
I_{v, e}=\left\{\begin{array}{ll}
1 & \text {if the edge } e \text { ends at vertex } v, \\
-1 & \text {if the edge } e \text { starts at vertex } v, \\
0 & \text { otherwise. }
\end{array}\right.
\end{align}

We then define two graph Laplacians
\begin{align}
&\Delta_{\Gamma} = II^{t} = D-A\, \label{eq:scalarLaplacian},\\ 
&\Delta_{E} = I^{t}I\,\label{eq:EdgeLaplacian}\, . 
\end{align}
Equation \eqref{eq:scalarLaplacian} represents the ordinary graph Laplacian $\Delta_\Gamma$, which is a $|V|\times|V|$ matrix. We will need the operator $\Delta_E$ in \eqref{eq:EdgeLaplacian} later when we discuss Fermions in Section \ref{sec:FreeFermionsonGraphs}. The degree matrix $D$ is diagonal with diagonal elements $D_{v,v} = d(v)$, which is the degree of vertex $v$, i.e., the number of edges connected to vertex $v$. The adjacency matrix $A$ is  defined as follows 
\begin{align}
\label{eq:adj}
A_{v, w}=\left\{\begin{array}{ll}
1 & \text{ if } (v,w) \text{ or } (w,v) \in E(\Gamma)\, , \\
0 & \text { otherwise. }
\end{array}\right.
\end{align}
Here one can interpret the matrix element $\left(\frac{1}{D}A\right)^{\t}_{v\,w}$ as the transition probability that a walk starts from a vertex $v$ and ends at a vertex $w$ in $\t$ steps.\footnote{A walk is a sequence of vertices that a walker passes through. A path is a walk without vertices repeated. Later, we will generalize the notion of these walks and paths to accommodate a combinatorial interpretation of Dirac operators and include edges in addition to vertices.} The matrices $D$ and $A$ are independent of the orientations of the edges, and therefore are properties of the underlying undirected graph. In Section \ref{sec:SFTonGraphs}, we will use these graph operators to construct scalar QFT on graphs.

We consider the space of functions on vertices such that, 
\be
 f_V : V(\Gamma) \to \mathbb{R}\, ,
\ee
and the space of anti-symmetric functions on edges
\be
f_E : V(\Gamma) \times V(\Gamma) \to \mathbb{R}\, ,
\ee
such that
\be
f_E (v,w) =
\begin{dcases}
    -f_E(w,v) &  \text{if}\, (v,w)\, \text{or}\,(w,v) \in E(\Gamma)\, , \\\
    0 & \text{otherwise} \, .
\end{dcases}
\ee
In the language of algebraic topology, the functions $f_V$ and $f_E$ form the $0$- and $1$- cochains, respectively. Cochains are a discrete notion of differential forms. The incidence matrix $I : f_V \to f_E $ works as a discrete notion of differential operator and the transpose $I^{t} : f_E \to f_V$  acts as a discrete co-differential operator. The combinatorial graph Laplacian $\Delta_{\Gamma}: f_V \to f_V$ describes the diffusion process on vertices through edges, and the edge Laplacian $\Delta_E : f_E \to f_E$ describes a diffusion process on edges through vertices \cite{Bianconi_2021}. The operator $\Delta := \Delta_{\Gamma} \bigoplus \Delta_E: f_V \bigoplus f_E \to f_V \bigoplus f_E $ is a discretization of the Hodge-Laplacian restricted to the first two cochain complexes. See \cite{Becher:1982ud,lim2019hodge} for the general expression applied to higher form cell complexes. 

 We define a scalar field on graphs as the map $\phi(x): V(\Gamma) \to \mathbb{R} $ where $x \in V(\Gamma)$. One can interpret $\phi(x)$ as the $x^{\text{th}}$ element of a $|V|$ sized vector. As a notational remark, any operator element $O(x,y)$ can also be  interpreted  as the $(x,y)$ element of the corresponding matrix.
 One now writes down a free scalar action
\begin{equation}
\label{eq:ScalarActionGraph}
S_{\Gamma}[\phi] = \sum_{x,y}\phi(x)C^{-1}(x,y)\phi(y) \,.
\end{equation}
Here, $C^{-1}(x,y) = \Delta_{\Gamma}(x,y) + m \mathbf{1}(x,y)$.
We define our partition function as 
\begin{equation}
\label{eq:ScalarPartitionfunction}
\mathcal{Z}[J]= \int \mathcal{D}\phi\exp \left[-S_{\Gamma}[\phi]+ \sum_{x}J(x)\phi_{x}\right]\, .
\end{equation}
By standard QFT techniques, we get the two-point function $C(x,y)$  as
\begin{equation}
\frac{1}{\mathcal{Z}[0]}\frac{\d^2 \mathcal{Z}[J]}{\d J(x) \d J(y)}\bigg\vert_{J=0} = \left(\frac{1}{\DG+m\mathbf{1}}\right)(x,y) \, ,
\end{equation}
where $\mathbf{1}$ is the $|V|\times|V|$ identity matrix. In  matrix notation, this is equivalent to the resolvent or the Green function $G(m)$ of $-\Delta_{\Gamma}$  with matrix elements $G(x,y;m) = C(x,y)$. One then sees that
\begin{equation}
\label{eq:resolventgeneralscalar}
G(m) = \frac{1}{\DG+m \mathbf{1}} = \frac{1}{D-A+m\mathbf{1}} = \frac{1}{D+m\mathbf{1}}\sum^{\infty}_{\t=0}\left(\frac{1}{D+m\mathbf{1}}A\right)^{\t} \, .
\end{equation}
A particular matrix element $G(x,y;m)$ can be interpreted as the transition probability (slightly enhanced or reduced by the term $m$ depending on the sign) of a walk starting from a vertex $x$ and ending at $y$ in all possible steps multiplied by a prefactor $\frac{1}{d(x)+m}$.\footnote{Here, $d(x) = D(x,x)$ is the degree of the vertex $x$.}  
$\left(\frac{1}{d}A\right)^{\t}(x,y)$ gives the transition probability of a simple random walk starting from $x$ and ending on $y$ \cite{norris1998markov}.
The last equality in \eqref{eq:resolventgeneralscalar}, for large enough mass, is a discrete version of Schwinger proper time parametrization of the two-point function (cf. \cite{gurau2014renormalization}).
Thus, the problem of computing $G(x,y;m)$ can be mapped to the problem of calculating random walk statistics on graphs.

\subsection{Free Fermions}
\label{sec:FreeFermionsonGraphs}

\subsubsection{Free Fermions in Euclidean spaces}
\label{subsec:FFeuclid}

For Fermions, we define two independent Grassmann fields  $\psi(x)$ and $\bar{\psi}(x)$ and introduce two local anticommuting sources $\eta(x)$ and $\bar{\eta}(x)$ where $x \in \mathbb{R}^{n}$. We proceed to define the free Dirac action
\begin{equation}
S[\psi, \bar{\psi}] 
= 
\int \dd[n] x \, \dd[n] y  \,
\bar{\psi}(x)C^{-1}(x,y)\psi(y) \,,
\end{equation}
and the partition function with sources
\beq
\mathcal{Z}[\eta, \bar{\eta}] 
= 
\int \mathcal{D}\psi\mathcal{D}\bar{ \psi} 
\, 
e^{-S[\psi, \bar{\psi}] +\bar{\eta}(x) \psi(x)+ \bar{\psi}(x) \eta(x)}\,.
\eeq
In the case of Fermions, $C^{-1}(x,y) = (\slashed{\p} + m )\d^{(n)}(x-y)$ is the Dirac operator, and 
$\slashed{\p} = \g_{\m}\p^{\m}$ (using Einstein summation convention). 
The $\gamma^{\m}$s are matrices that satisfy the Clifford algebra 
\beq
\{\gamma^{\m},\gamma^{\n}\}=g^{\m\n}\,,
\eeq
in $n$ dimensions, and gives us a spin representation of the $\text{SO}(n)$ group on $\mathbb{R}^{n}$. 
Here $g^{\m\n}$ is the Euclidean metric and $\{\cdot,\cdot\}$ represents the anti-commutator. 
One can deduce that the two-point function now in real space representation is given by
\beq
G_{\slashed{\p}}(x,y;m) = C(x,y) = \left[\frac{1}{\slashed{\p}+m}\right]\left(x,y\right)\,.
\eeq
A standard reference for Dirac operators and spinors on Riemannian geometry is \cite{friedrich2000dirac}.

\subsubsection{Free Fermions on graphs and the Dirac walk}
\label{sec:FermionicQFt}

Fermions have been well studied in discrete spaces in the context of lattice QFT \cite{Montvay:1994cy} where the underlying graph is $\mathbb{Z}^{n}$. 
The presence of $\gamma$ matrices in the Fermionic $C(x,y)$ makes its definition heavily reliant on the underlying $SO(n)$ symmetry of the background space. 
This particular symmetry is not enjoyed by more general discrete spaces. The problem is not so critical in $\mathbb{Z}^{n}$ as one still inherits the discrete rotational symmetries of $\mathbb{R}^{n}$ but in the case of a general graph this may not be true.

In this section, we propose a possible definition of a Fermionic QFT on graphs without alluding to $\gamma$ matrices. 
We observe that 

\beq
\label{eq:gammaMatAndLaplacian}
(\slashed{\p}+m)(m-\slashed{\p}) = -g^{\m\n}\p_{\m}\p_{\n}+ m^{2} = -\Delta_{\text{scalar}} + m^2 \,,
\eeq
is nothing but the Laplacian acting on scalars in $\mathbb{R}^{n}$. Thus $\slashed{\p}$ is a formal square root of the Laplacian.

We look at the geometric notion of Dirac equation in continuum (without referring to $\gamma$ matrices), first introduced by K\"ahler \cite{Kahler1962}, known as the K\"ahler-Dirac equation 
\be
\left(d-\delta+m_0\right) \Psi(x)\, =0.
\ee
In $n$ dimensions, the K\"ahler-Dirac Fermion is described as a linear combination of differential forms (rather than as spinors)
\be
\Psi(x)=\stackrel{\circ}{\psi}(x)+\psi_{\mu_{1}}(x) d x^{\mu_1}+\frac{1}{2 !} \psi_{\mu_1 \mu_2}(x) d x^{\mu_1} \wedge d x^{\mu_2} + \ldots +  \psi_{\mu_1 \ldots \mu_n}(x) dx^{\mu_1}\wedge\ldots dx^{\mu_n}\, ,
\ee
where the index $\mu_i$ ranges from $1$ to $n$, $d$ is the differential operator or the exterior derivative, $\delta$ is the co-differential operator, and $(d - \delta)$ is the K\"ahler-Dirac operator.
The square of the K\"ahler-Dirac operator
\be
(d-\delta)^2 = -(d\delta + \delta d) = -\Delta_{\text{Hodge}} \, ,
\ee
is the $n$-dimensional Hodge-Laplacian acting on the space of differential forms, thus the K\"ahler-Dirac operator is the formal square root of the Hodge-Laplacian similar to the usual Dirac operator $\slashed{\partial}$. In fact, the K\"ahler-Dirac equation can be shown to be equivalent to the Dirac equation in $\mathbb{R}^n$ \cite{Becher:1982ud}.

For path-integral quantization of K\"ahler-Dirac fields we consider the following action \cite{Becher:1982ud}
\be
S_{KD}=\int \left(\bar{\Psi},\left( d -\delta+m_0\right) \Psi\right)_{0}\,.
\ee
Here $\Psi$ is a collection of Grassmann valued $p$-form fields, and $\bar{\Psi}$ is a similar independent Grassmann valued field.
The notation $\left(\cdot,\cdot\right)_{0}$ is a scalar product in the space of differential forms as defined in \cite{Becher:1982ud,Kahler1962}.

Taking motivation from the K\"ahler-Dirac operator in the continuum, we define a graph Dirac operator \cite{Cassidy:https://doi.org/10.48550/arxiv.2203.02782,Knill:https://doi.org/10.48550/arxiv.1306.2166} as a $(|V|+|E|)$ squared matrix (
 $\sD: f_V \bigoplus f_E \to f_V \bigoplus f_E$)

\beq
\label{eq:SlashedD}
\slashed{D}=\left(\begin{array}{ll}
0 & I \\
I^{t} & 0
\end{array}\right) \,,
\eeq
and we can see 
\beq
\slashed{D}^{2}=\left(\begin{array}{cc}
\Delta_{\Gamma} & 0 \\
0 & \Delta_{E}
\end{array}\right)\,,
\eeq
is the formal square root of the generalized Hodge-Laplacian matrix.

We consider the discrete version of K\"ahler-Dirac fermion with a Grassmann-valued cochain complex $\Phi= \psi_V \bigoplus \chi_E$ as the Fermionic field \cite{Becher:1982ud, catterall2018kahler,Bianconi_2021,Bianconi_2023,Bianconi_2024}. We restrict ourselves to spaces of up to $1$- cochains as our computations in sections \ref{ch-2} and \ref{ch-5} are on tree graphs where higher cochains do not exist. As discussed in Section \ref{sec:SFTonGraphs}, cochains are discrete versions of differential forms. Here, $\psi_V(v)$ is a Grassmann valued $0$-cochain localized on the vertex and $\chi_E(e)$ is a Grassmann valued $1$-cochain localized on edges. Practically, we can view $\Phi = \left[\begin{array}{l}\psi_V \\ \chi_E\end{array}\right]$, a column matrix of dimension $1 \times (|V|+|E|)$. Similarly we write another independent Grassmann valued field $\bar{\Phi}=\left[\begin{array}{ll}
\bar{\psi}_V & \bar{\chi}_E
\end{array}\right]$, a chain complex.
\footnote{The bar is only a formal notation and doesn't refer to transposition and complex conjugation.}

Edges and vertices form the set of sites and the distance between two sites $x$ and $y$ is the number of half-edges of the geodesic (i.e. the shortest path) $g(x,y)$ between the two sites, noted $\abs{g(x,y)}$.

For the functional integral, we define Grassmann integral on vertices and edges in the standard way \cite{Peskin:1995ev},and the path-integral measure as follows
\be
\mathcal{D}\bar{\Phi}\mathcal{D}\Phi  = 
\prod_{v \in V(\Gamma)}d \bar{\psi}_V(v)\,d \psi_V(v) 
\prod_{e \in E(\Gamma)}d \bar{\chi}_E(e)\,d \chi_E(e) \,.
\ee
Consequently,
for any $(|V|+|E|)$-square matrix  $B$, the Gaussian integration gives
\be
\int \mathcal{D} \bar{\Phi}\mathcal{D}\Phi\,\, e^{-\bar{\Phi} B \Phi} = \operatorname{det}(B)\, .
\ee

This allows us to write the action, partition function  and correlation function using the usual Gaussian integration techniques 
\al{
\label{eq:DiracGraphAction}
&S[\Phi,\bar{\Phi}] = -\left[\bar{\Phi}( -\slashed{D} )\Phi+m \bar{\Phi} \mathbf{1} \Phi\right]\,,\\ \label{eq:DiracGraphPartition}
&\mathcal{Z}[ \eta, \bar{\eta}]=\int \mathcal{D} \bar{\Phi} \mathcal{D} \Phi e^{-(\left[\bar{\Phi}( -\slashed{D} )\Phi+m \bar{\Phi} \mathbf{1} \Phi\right]
+\bar{\eta} \Phi+\bar{\Phi} \eta)}\,,\\
&\left\langle\bar{\Phi}(x) \Phi(y)\right\rangle=C(x,y)=\left(\frac{1}{-\slashed{D}+m \mathbf{1}}\right)(x,y)=\frac{1}{m} \sum_{\t=0}^{\infty}\left(\frac{\slashed{D}}{m}\right)^{\t}(x,y)\,. \label{eq:ResolventgeneralDirac}
}
We recall that $x$ and $y$ refer to sites and $\Phi(x)$ will correspond to $\psi(x)$ if $x$ is a vertex or to $\chi(x)$ if $x$ labels an edge. 

\begin{figure}[H]
    \centering
    \begin{minipage}[t]{.75\textwidth}
    \centering
    \includegraphics[width = 7cm]{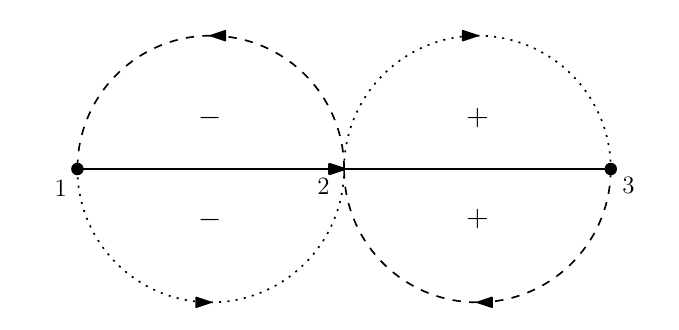}
    \caption{A walk going from vertex $1$ to edge $2$  or $2$ to $1$  will get a ``$-$'' sign because the edge $2$ points outwards from $1$. A walker moving from $2$ to $3$  or $3$ to $3$  will get a ``$+$" as the edge points towards $3$.
  }
    \label{fig:diracstep}
    \end{minipage}
\end{figure}

Like powers of the adjacency matrix $A$, the powers of the operator $\sD$ have a combinatorial interpretation in terms of signed walks on the graph, as alternating sequences of edges and vertices, which henceforth will be called `Dirac walks'. The set of all Dirac walks of starting point $x$, ending point $y$, and number of steps $\tau$ is denoted by $\Omega_{\t, x\to y}$, their cardinality by $\abs{\Omega_{\t, x\to y}}$, while a particular walk will be written as $\omega$. \footnote{To add to clarity, we may add a subscript to $\omega$ in order to specify some relevant features of the walk.}
A Dirac walk of one step $\omega_{\tau = 1}$ moves from a vertex to an incident edge or an edge to an incident vertex. We will define the sign of the walk $\omega_{\tau = 1}$ as introduced in \cite{Cassidy:https://doi.org/10.48550/arxiv.2203.02782}: if the edge involved in the walk is directed out of the vertex involved, then $\operatorname{sgn}(\omega_{\tau = 1})=-1$, and if the edge enters the vertex involved, then $\operatorname{sgn}(\omega_{\tau = 1})=1$ (see the Figure \ref{fig:diracstep}). 
The sign of a walk is defined as the product of all the signs of the individual steps taken.
The quantity $ \sD^{\t}(x, y)=\sum_{\omega \in \Omega_{\t, x \rightarrow y}} \text{sgn}(\omega)$ gives us the sum of signed walks between sites $x$ and $y$ in $\t$ steps. 

\begin{figure}[H]
    \centering
    \begin{minipage}[t]{.5\textwidth}
    \centering
    \includegraphics[width = 4cm]{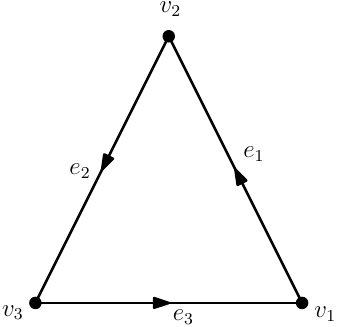}
    \caption{A directed complete graph $K_3$.}
    \label{fig:3cycle}
    \end{minipage}
\end{figure}

Analogously to Section \ref{sec:FermionicQFt}, we obtain a Schwinger proper time parametrization of our $C(x,y) = G_{\sD}(x,y;m)$, which allows us to compute $C(x,y)$ combinatorially.
We will demonstrate the properties of powers of $\sD$, introduced in Section \ref{sec:SFTonGraphs}, with an example of directed 3-cycle $K_3$ as shown in Figure \ref{fig:3cycle}. 

It has the following incidence matrix
\al{I=
\left(
\begin{array}{ccc}
 -1 & 0 & 1 \\
 1 & -1 & 0 \\
 0 & 1 & -1 \\ 
\end{array}
\right) \,,
}
and by definition, the graph Dirac operator is given by
\be
\slashed{D}=
\left(
\begin{array}{cccccc}
 0 & 0 & 0 & -1 & 0 & 1 \\
 0 & 0 & 0 & 1 & -1 & 0 \\
 0 & 0 & 0 & 0 & 1 & -1 \\
 -1 & 1 & 0 & 0 & 0 & 0 \\
 0 & -1 & 1 & 0 & 0 & 0 \\
 1 & 0 & -1 & 0 & 0 & 0 \\
\end{array}
\right)\,.
\ee
Its third power is
\be
\slashed{D}^{3}=\left(\begin{array}{cccccc}
0 & 0 & 0 & -3 & 0 & 3 \\
0 & 0 & 0 & 3 & -3 & 0 \\
0 & 0 & 0 & 0 & 3 & -3 \\
-3 & 3 & 0 & 0 & 0 & 0 \\
0 & -3 & 3 & 0 & 0 & 0 \\
3 & 0 & -3 & 0 & 0 & 0
\end{array}\right)
\,,
\ee
with the corresponding possible three-step walks starting from the vertex $v_1$
\al{
\omega_1=v_1 \rightarrow e_3 \rightarrow v_3 \rightarrow e_2 & \quad\operatorname{sgn}\left(\omega_1\right)=-1 \,,\nonumber\\
\omega_2=v_1 \rightarrow e_1 \rightarrow v_2 \rightarrow e_2 & \quad\operatorname{sgn}\left(\omega_2\right)=1 \,,\nonumber\\
\omega_3=v_1 \rightarrow e_3 \rightarrow v_3 \rightarrow e_3 & \quad\operatorname{sgn}\left(\omega_3\right)=1 \,,\nonumber\\
\omega_4=v_1 \rightarrow e_3 \rightarrow v_1 \rightarrow e_3 & \quad\operatorname{sgn}\left(\omega_4\right)=1 \,,\nonumber\\
\omega_5=v_1 \rightarrow e_1 \rightarrow v_1 \rightarrow e_3 & \quad\operatorname{sgn}\left(\omega_5\right)=1 \,,\nonumber\\
\omega_6=v_1 \rightarrow e_1 \rightarrow v_2 \rightarrow e_1 & \quad\operatorname{sgn}\left(\omega_6\right)=-1 \,,\nonumber\\
\omega_7=v_1 \rightarrow e_1 \rightarrow v_1 \rightarrow e_1 & \quad\operatorname{sgn}\left(\omega_7\right)=-1 \,,\nonumber\\
\omega_8=v_1 \rightarrow e_3 \rightarrow v_1 \rightarrow e_1 & \quad\operatorname{sgn}\left(\omega_8\right)=-1 .
}
There are two walks, $\omega_{1}$ and $\omega_{2}$, between $v_1$ and $e_2$ with opposite signs,
therefore the matrix element in $\sD^{3}$ corresponding to three-step walks between $v_1$ and $e_2$ is zero.

\section{Walks on the Bethe lattice}
\label{ch-2}

Let us start this section by introducing the graph of our concern. The Bethe lattice is a $d$-regular infinite tree, i.e., each vertex in the tree is connected to $d$ other vertices. We are interested in counting Dirac walks and simple random walks between two lattice sites on the Bethe lattice. When we study simple random walks, the lattice sites refer to vertices of the graph; and when we discuss Dirac walks, lattice sites refer to both vertices and the directed edges of the underlying graph.

In this Section \ref{ch-2}, we will start by reviewing well known results of simple random walks on the Bethe lattice. At first, we will study the $d=2$ Bethe lattice which is simply the integer line $\mathbb{Z}$. We will then present the results for $d \geq 3$ Bethe lattice by mapping the Bethe lattice to a half-line as discussed in \cite{Monthus_1996,Hughes1982}.
Based on the generating function techniques used by \cite{Monthus_1996,Hughes1982}, we will adapt our technique to understand the Dirac walk and present new results on its combinatorics on the Bethe lattice.

\subsection{Review of the simple random walk}
{\label{sec:Walkonline}}

\subsubsection{Simple random walk on the line}
\label{sec:RWalkonline}

\begin{figure}[H]
    \centering
    \begin{minipage}[t]{.75\textwidth}
    \centering
    \includegraphics[width = 12cm]{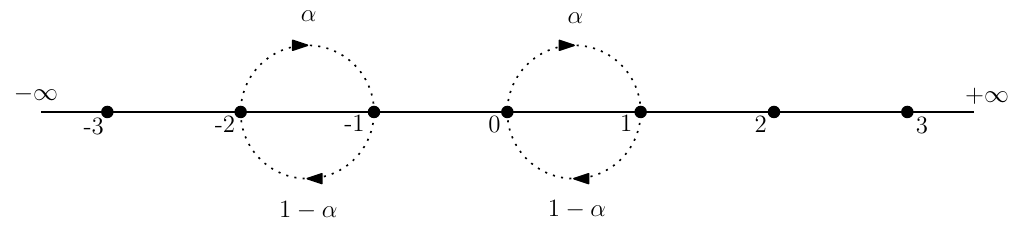}
    \caption{A section of the integer line $\mathbb{Z}$. The probability of moving towards $+\infty$ is $\a$ (e.g. $-2 \to -1$ and $0 \to 1$) and probability of moving towards $-\infty$ is $1-\a$ (e.g. $-1 \to -2$ and $1 \to 0$).}
    \label{fig:lineSRW}
    \end{minipage}
\end{figure}

For understanding simple random walks on a line, we choose a coordinate system in such a way that vertices are labeled by some $x \in \mathbb{Z}$, and $x=0$ is referred to as the origin.
For our system of focus, let the probability of a random walker going from some vertex $x$ to $x+1$ in a single step be $\a$ and going from $x$ to $x-1$ in one step be $1-\a$ (cf. Figure~\ref{fig:lineSRW}). 
\noindent Here, $\a$ is a real number such that $0< \a < 1$. 
We define $h_{\t}(0,x)$ as the probability that a random walker is at some lattice site $x$ in $\t$ steps where $\t$ is a non-negative integer, starting from some origin $0$; therefore $0 \leq h_{\t}(0,x) \leq 1$. The quantity $h_{\t}(0,x)$ follows the recursion
\begin{align} \label{eq:RecursionforRWline}
h_{\t+1}(0,x) = \a\,h_{\t}(0,x-1)+ (1-\a)\,h_{\t}(0,x+1)
\,,
\end{align}
where the boundary condition reads
\beq
h_{0}(0,x) = \delta_{x,0}\,.
\eeq
We define the generating function of the transition probability as
\beq 
\label{eq:Generating Function}
H_{0,x}(\l) = \sum_{\t=0}^{\infty} \l^{\t}h_{\t}(0,x)\,.
\eeq
This generating function will become the cornerstone of our analysis of the Green function. We multiply Equation \eqref{eq:RecursionforRWline} by $\l^{\t+1}$ and perform a summation to get
\begin{align}
\sum_{\t=0}^{\infty}\l^{\t+1}h_{\t+1}(0,x) &= \l \sum_{\t=0}^{\infty}\l^{\t}\left( \a\,h_{\t}(0,x-1) + (1-\a)\,h_{\t}(0,x+1)\right)\,, \nnn
&= H_{0,x}(\l) - \delta_{x,0} = \l\left(\a\,H_{0,x-1}(\l)+ (1-\a)\,H_{0,x+1} (\l)\right)
\,.
\label{eq:Recursionfor1dRW}
\end{align}
Owing to the linearity and homogeneity in $H_{0,x}$ of \eqref{eq:Recursionfor1dRW} when $x$ is non-zero, we can define two index shifting operators, $T$ and $S$, such that 
\be
T H_{0,x}(\l) = H_{0,x+1}(\l) 
\,,
\iif x \geq 0 
\,, 
\label{eq:Upshift1}
\ee
and 
\be
S H_{0,x}(\l) = H_{0,x-1}(\l) 
\,,
\iif x \leq 0
\,.
\label{eq:Downshift1}
\ee
When $x \geq 1$, the recursion \eqref{eq:Recursionfor1dRW} can be rewritten as
\be \label{eq:ShitedRecursion}
TH_{0,x-1}(\l) = \l\left(\a\,H_{0,x-1}(\l)+ (1-\a)T^{2}\,H_{0,x-1} (\l)\right)
\,.
\ee
The condition $x\geq 1$ ensures that by applying the operator $T$, the L.H.S. of \eqref{eq:Recursionfor1dRW} does not reach $H_{0,0}(\l)$, as the recursion has a different form for the latter.
As $\a$ is independent of $x$, we can factorize \eqref{eq:ShitedRecursion} and get 
\be
\big(T - t_{+}(\l)\big)\big(T-t_{-}(\l)\big)H_{0,x-1}(\l) = 0 \, .
\ee
$T - \l \left( \a + (1-\a)T^{2}\right) = 0$ is the characteristic equation and $t_{+}(\l)$ and $t_{-}(\l)$ are called the characteristic roots. The roots are written as follows
\be
t_{\pm}(\l) = \frac{1\pm\sqrt{1-4 (1-\alpha) \alpha  \lambda ^2}}{2 (1-\a) \lambda }
\,.
\ee
The recursion \ref{eq:Recursionfor1dRW} now boils down to two independent linear recursions when $x \geq 1$.
\be
(T- t_{\pm}(\l))H_{0,x-1}(\l)= 0\, .
\ee
We pick the linear recursion with $t_{-}(\l)$ which is regular for $\l \to 0$. We set $t_{-}(\l) = t(\l)$. Therefore,
\be
T H_{0,x-1}(\l) = H_{0,x}(\l)= t(\l)H_{0,x-1}(\l) \, ,
\iif x \ge 1
\ee
which  leads to
\be
\label{eq:recRightSide}
H_{0,x}(\l) =(t(\l))^{x} H_{0,0}(\l) 
\,,
\quad \text{if}\, x \geq 0
\,.
\ee
A similar exercise can be performed when $x \leq -1$ and we obtain the characteristic equation
\be
\Big(S - \l\left(\a S ^{2} + (1-\a)\right)\Big)H_{0,x+1}(\l) = 0  \, ,
\ee
with characteristic roots
\be
s_{\pm}(\l) = \frac{1\pm\sqrt{1-4 (1-\alpha) \alpha  \lambda ^2}}{2 \alpha  \lambda } \, .
\ee
We similarly pick up the characteristic root $s_{-}(\l)$, which is regular for $\l \to 0$ and set $s_{-}(\l) = s(\l)$,
\be
S H_{0,x}(\l) = H_{0,x-1}(\l) = s(\l)H_{0,x}(\l)\, ,
\ee
and therefore,
\be
\label{eq:recLeftSide}
H_{0,x}(\l) = (s(\l))^{|x|}H_{0,0}(\l)\,, \quad \text{if }\,x \leq 0 \, . 
\ee
For $x=0$, Eq. \eqref{eq:Recursionfor1dRW} becomes
\be
H_{0,0}(\l)-1 = \l(\a\,H_{0,-1}(\l)\,+ (1-\a)\,H_{0,1}(\l))\,.
\ee
By substituting $H_{0,1}(\l) = H_{0,0}(\l)t(\l)$ and $H_{0,-1}(\l) = H_{0,0}( \l) s(\l)$ we get
\begin{align}
H_{0,0}(\l)-1 &= \l \big(\a\,H_{0,0}(\l)\,s(\l)+ (1-\a)\,t(\l)H_{0,0}(\l)\big)\, ,\\
H_{0,0}(\l) &= \frac{1}{\sqrt{1-4 (1-\alpha) \alpha  \lambda ^2}}\, ,
\end{align}
implying from \eqref{eq:recRightSide} and \eqref{eq:recLeftSide},
\be
 H_{0,x}(\l) =
\begin{dcases}
\frac{
\left(
\frac{
1-
\sqrt{
1-4 (1-\a) \alpha  \lambda ^2}
}
{
2 (1-\a) 
\lambda }
\right)
^{x}
}
{
\sqrt{
1-4 (1-\alpha) \alpha  \lambda ^2
}
} 
\,,
&\quad \text{if}\, x\geq 0\, ,\\
\frac{
\left(
\frac{
1-
\sqrt{
1-4 (1-\alpha) \alpha  \lambda ^2
}
}
{
2 \alpha  \lambda 
}
\right)
^{|x|}}
{
\sqrt{
1-4 (1-\alpha) \alpha  \lambda ^2
}
}
\,,
&\quad \text{if}\, x\leq 0\,.
\end{dcases}
\ee
For $\a = \frac{1}{2}$, we have $s(\l) = t(\l)$,
and we get
\be
H_{0,x}(\l) = H_{0,0}(\l) (t(\l))^{|x|}\, ,
\ee
and 
\be
H_{0,x}(\l) = \frac{\left(\frac{1-\sqrt{1-\lambda ^2}}{\lambda }\right)^{|x|}}{\sqrt{1-\lambda ^2}}\, .\label{eq:GFforlineSRW0}
\ee 
From Eq. \eqref{eq:GFforlineSRW0}, we can derive the transience or the recurrence of a random walk. A random walk is called transient if
\beq
\sum_{\t=0}^{\infty} h_{\t}(0,0) 
= \lim_{\l  \to 1}H_{0,0}(\l) < \infty\,,
\eeq
otherwise, they are called recurrent \cite{norris1998markov}. 
For the case of biased random walk ($\a \neq \f{1}{2}$), we have
\be
\lim_{\l  \to 1}
H_{0,0}(\l)
= 
\frac{
1
}
{
\sqrt{1-4 (1-\alpha) \alpha
}
}
\, ,
\ee
and for an unbiased random walk ($\a = \frac{1}{2}$)
\be
\lim_{\l \to 1}
H_{0,0}(\l) 
\to
\infty
\, .
\ee
Therefore, the unbiased simple random walk in one dimension is recurrent, whereas one can easily read off that the biased random walk is transient.

\subsubsection{Simple random walk on the $d \geq 3$ Bethe lattice} \label{RandomwalkBethe}

\begin{figure}[H]
\centering
\begin{minipage}[t]{.75\textwidth}
\centering
\includegraphics[width=6.5cm]{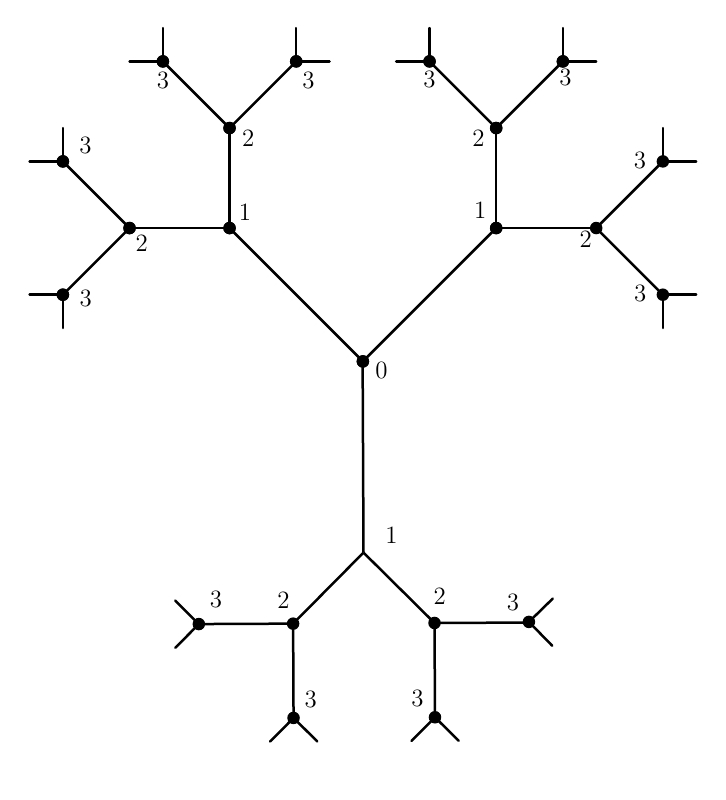}
\caption{
A $d =3$ Bethe lattice. The vertex labels give their distance to the origin of the lattice in units of edge length.
}
\label{fakelebal1}
\end{minipage}
\hfill
\centering
\begin{minipage}[b]{.75\textwidth}
\centering
\includegraphics[width=7.5cm]{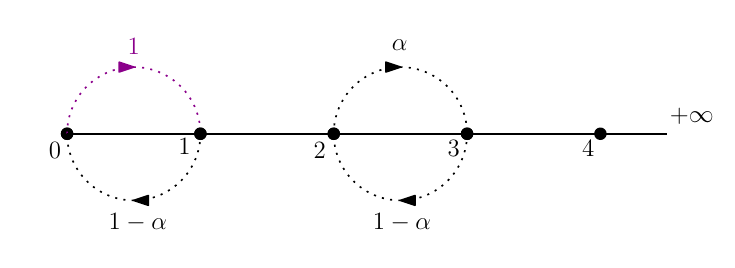}
\caption{
The half-line to which the Bethe lattice is mapped. The (anomalous) weight 1 on the step between vertices $0$ and $1$ reflects the boundary conditions \eqref{eq:bdycondhalfline}.
}
\label{fig:fake}
\end{minipage}
\label{fig:3-betherlattice}
\end{figure}

Random walks on $d\geq 3$ Bethe lattice have been extensively studied in \cite{Hughes1982,Monthus_1996} and references within. 
They used a coordinate system where some vertex is chosen as a origin and is labeled $0$. As all the vertices are topologically equivalent, one is free to choose any vertex as the origin. All the vertices which are $x$ edges away from the origin are labeled $x$, where $x$ is a non-negative integer. There are $d(d-1)^{x-1}$ such vertices which are $x$ edges away from the origin. To compute the simple random walk statistics, one then maps all the vertices $x$ edges away from the origin of the Bethe lattice to a vertex $x$ edges away from the origin of the half-line. A visual depiction of a Bethe lattice can be found in Figure \ref{fakelebal1} and corresponding mapping to the halfline in Figure~\ref{fig:fake}. 

If a walk is on a vertex at $x \neq 0$ edges away, then in one step, the probability of going one step further away from $0$ is $\f{d-1}{d}$ and to go to one step closer to $0$ is $\f{1}{d}$. If a simple random walker is at $x=0$, it ends up at $x=1$ in one step with probability $1$. 
This observation allows us to map the simple random walk on the Bethe lattice to a random walk on the half-line with drift at the cost of losing the information about the angular degrees of freedom. 
The mapping is obtained through the recursion \eqref{eq:RecursionforRWline} with $\a = \f{d-1}{d}$. 
For the moment, we will perform the calculation for general $\a$ and then substitute $\a = \frac{d-1}{d}$ for the final result. The recursion for a simple random walk on the half-line  with a  drift starting from $0$ is, when $x \geq 2$,
\beq
\label{eq:RecursionforRWbethe}
h_{\t+1}(0,x) = \a h_{\t}(0,x-1) + (1-\a) h_{\t}(0,x+1)\,,
\eeq
and with  reflecting boundary conditions at $0$,
\begin{align}
\label{eq:bdycondhalfline}
h_{0}(0,x) &= \delta_{x,0} \,, \\
h_{\t+1}(0,0) &= (1-\a)h_{\t}(0,1)\,, \\
h_{\t+1}(0,1) &= h_{\t}(0,0) +  (1-\a) h_{\t}(0,2)\,.  
\end{align}
The prefactor for $h_{\t}(0,1)$ is $1$ as we will surely go to lattice sites labeled $1$ in the $(\t+1)$-th step if we are at $0$ after $\t$ steps.
These equations can be solved with the generating function methods introduced in the Section \ref{sec:Walkonline}.
The resulting generating functions concerning the boundary are 
\begin{align}
H_{0,0}(\l)-1
&= 
\l (1-\alpha)H_{0,1}(\l) 
\,,
\label{eq:BoundaryRW1} 
\\ 
H_{0,1}(\l) 
&= 
\l \left( H_{0,0} (\l)+ (1-\a)H_{0,2}(\l)  \right),
\label{eq:BoundaryRW2}
\end{align}
and for the interior ($x\geq 2$)  is
\be{
\label{eq:BetheBulkRW}}
H_{0,x}(\l) 
= 
\l\left(\a\,H_{0,x-1}(\l)+ (1-\a)\,H_{0,x+1} (\l)\right) 
\,.
\ee
Similarly to \eqref{eq:Recursionfor1dRW}, for $x\geq2$, Equation \eqref{eq:BetheBulkRW} can be written in terms of the index shift operator $T$, defined in \eqref{eq:Upshift1}.
This leads to same characteristic equation for the positive sector of \eqref{eq:Recursionfor1dRW}, with different boundary conditions with the solution
\be
\label{eq:linearrecRW}
T H_{0,x-1}(\l) = H_{0,x}(\l) 
= 
t(\l)H_{0,x-1}(\l)
\,,
\quad \text{if } x \geq 2 \, ,
\ee
and thus,
\be
\label{eq:H0xt}
H_{0,x}(\l) 
= 
t(\l)
^{x-1}
H_{0,1}(\l)
\,,
\iif
x \ge 1
\, .
\ee
We then use $H_{0,2}(\l)= H_{0,1}(\l)t(\l)$ to solve for $H_{0,1}(\l)$ and $H_{0,0}(\l)$ for Equations \eqref{eq:BoundaryRW1} and \eqref{eq:BoundaryRW2}.
After substitution, one finds 
\be
H_{0,0}(\l) = \frac{2 \alpha }{\sqrt{1-4 (1-\alpha) \alpha  \lambda ^2}+2 \alpha -1}
\,,
\ee
and 
\be
H_{0,1}(\l) = \frac{2 \lambda }{1-2 (1-\alpha) \lambda ^2+\sqrt{1-4 (1-\alpha) \alpha  \lambda ^2}}
\,.
\ee
Going further  using \eqref{eq:H0xt},
\be
H_{0,x}(\l) = \frac{2 \left(\frac{1-\sqrt{1-4 (1-\alpha) \alpha  \lambda ^2}}{2 (1-\alpha) \lambda }\right)^x}{\sqrt{1-4 (1-\alpha) \alpha  \lambda ^2}+2 \alpha -1}\,,
\quad\,\text{if}\,x \geq 1\,.
\ee
Now we are in a position to replace $\a = \f{d-1}{d}$. We get 
\begin{align}
H_{0,0}(\l) 
&=
\frac{2 (d-1)}{\sqrt{d^2-4 (d-1) \lambda ^2}+d-2}
\,, 
\\
H_{0,x}(\l) 
&= 
\frac{2 d \left(\frac{d-\sqrt{d^2-4 (d-1) \lambda ^2}}{2 \lambda }\right)^x}{\sqrt{d^2-4 (d-1) \lambda ^2}+d-2} 
\,,
\quad\,\text{if}\,x \geq 1\,.
\end{align}
We are interested in computing the probability of reaching one particular vertex at $x$ edges away on the Bethe lattice. 
We can then, by symmetry, extract the information of probability of reaching a particular site on the Bethe lattice at $x$ by computing $\hat{H}_{0,x}(\l) := \frac{H_{0,x}(\lambda)}{d(d-1)^{x-1}}$, if $x \ge 1$ and $\hat{H}_{0,0}(\l) := H_{0,0}(\l)$,
\beq
\label{eq:singlesiteBetheRW}
\hat{H}_{0,x}(\l) = \frac{2 (d-1) \left(\frac{d-\sqrt{d^2-4 (d-1) \lambda ^2}}{2 (d-1) \lambda }\right)^x}{\sqrt{d^2-4 (d-1) \lambda ^2}+d-2}\,.
\eeq
Remark that $\hat{H}_{x,y}(\l)$ is the generating function for the quantity $\hat{h}_{\t}(x,y)$, which will be introduce later in Section \ref{sec:originshift} as the transition probability that a random walker starts on a particular vertex $x$ and ends at a particular vertex $y$ in $\t$ steps on a Bethe lattice. 
In the case of the Bethe lattice, we observe that 
\be
\lim_{\l \to 1}
\hat{H}_{0,0}(\l) 
= 
\f{
d-1
}
{
d-2
}
<
\infty\,,
\ee
implying that the simple random walk on the Bethe lattice  for $d \ge 3$ is transient.

\subsection{Dirac walk}
\label{sec:Diracwalkgeneral}
\subsubsection{Dirac walk  on the line}
\label{ssec:DiracWalkonline}

\begin{figure}[H]
    \centering
    \begin{minipage}[t]{.8\textwidth}
    \centering
    \includegraphics[width=13cm]{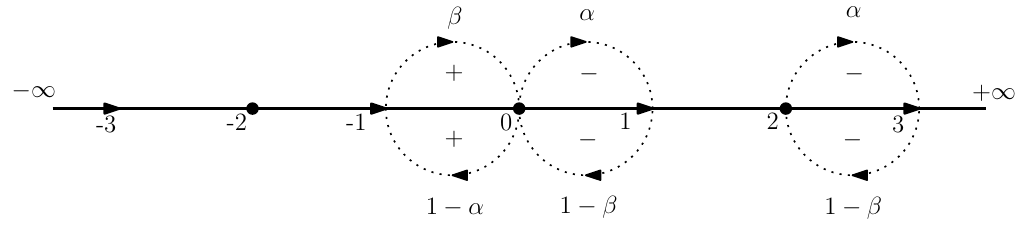}
    \caption{A directed line graph with all edges pointing from $-\infty$ to $\infty$ and the associated weights for a single step of a Dirac walk, from vertex to edge and edge to vertex.
    }
    \label{fig:directedline}
    \end{minipage}
\end{figure}

Based on the techniques reviewed in section \ref{sec:RWalkonline}, 
we want to present some new results on the combinatorics of the process generated by the Dirac operator, the Dirac walk, on the Bethe lattice. We assign a coordinate system on the $d=2$ Bethe lattice such that a vertex is labeled as the origin, and the edge incident to the right of the origin is labeled $1$, and the edge incident to the left of $0$ is $-1$. 
Furthermore, every edge is labeled by an odd number $2x+1$ and every vertex is labeled by an even number $2x$, where $x \in \mathbb{Z}$. We will assume that all the edges are oriented from $-\infty$ to $+\infty$ as shown in Figure \ref{fig:directedline}. 

We introduce $\alpha$ (resp. $1-\alpha$) as the magnitude of the weight of the Dirac walk from a vertex to an edge towards $+\infty$ (resp. $-\infty$). 
Similarly, we introduce $\beta$ (resp. $1-\beta$) as the magnitude of the weight of the Dirac walk from an edge to a vertex towards $+\infty$ (resp. $-\infty$) direction.
If the edge of concern is incoming (resp. outgoing) with respect to the vertex of concern, then we furthermore assign $+$ (resp. $-$) to the weight as in Figure \ref{fig:diracstep}.
As a result, on the $d=2$ Bethe lattice with all the orientations of all the edges directed from $-\infty$ to $+\infty$, the weight of a Dirac walk starting from $2x$ to the neighboring edge $2x+ 1$ (resp. $2x-1$) is $-\alpha$ (resp. $+(1-\alpha)$).
Similarly, we assign the weight $+\beta$ (resp. $-(1-\beta)$) for the Dirac walk starting from $2x+1$ to $2x+2$ (resp. $2x$). 
See Figure \ref{fig:directedline}.

We define $p_{\t}(z,2x)$ to be the weight of a walk starting from some lattice site $z \in \mathbb{Z}$ and arriving at any vertex $2x$ in $\t$ steps, and similarly, $f_{\t}(z,2x+1)$ as the weight of a walker arriving at any edge $2x+1$ in $\t$ steps. 
The corresponding generating functions are 
\begin{gather}
P_{z,2x}(\l) = \sum_{\t=0}^{\infty} \l^{\t}p_{\t}(z,2x)\,,\\
F_{z,2x+1}(\l) = \sum_{\t=0}^{\infty} \l^{\t}f_{\t}(z,2x+1)\,.
\end{gather}

In the following computation we will consider the walker starting from the vertex $ z = 0$. This implies that we will be on a vertex in even number of steps and on an edge in odd number of steps.

The weights of the possible steps do not add up to one owing to the signed nature of the walk, so the Dirac walk is not probabilistic. Nevertheless, the weights are defined in such a way that, if we multiply by the total number of walks in $2\t$ steps, i.e., $2^{2\t}$ and compute $2^{2\t}p_{2\t}(0,2x)$, we will get the signed sum of Dirac walks from $0$ to $2x$ in $2\t$ steps, i.e $\sD^{2\t}(0,2x)$. This definition of weights will be instrumental in the discussion of Green function \eqref{eq:ResolventgeneralDirac} in Section \ref{ch-5}.
The weights are related by the following recursion for any $x$,
\begin{align}
\label{eq:Recursioninstepandposa}
&f_{2\t+1}(0,2 x+1)=(1-\alpha )\,p_{2 \tau} (0,2 x+2)-\alpha\,p_{2 \tau}(0,2 x)  
\,,
\\
&
p_{2 \tau +2}(0,2x)=\beta \,f_{2 \tau +1}(0,2 x-1)-(1-\beta )\,f_{2 \tau +1}(0,2 x+1)\,
\end{align}
and the boundary condition
\beq
p_{0}(0,2x) = \delta_{2x,0}\,.
\eeq
At the level of generating functions, the recursions turn out to be, for any $x$,
\begin{align}
\label{eq:Dirac1dRecursion1}
F_{0,2 x+1}(\l)&
= 
\l\big((1-\alpha)\,P_{0,2 x+2}(\l)-\alpha\,P_{0,2 x}(\l)\big)
\,, \\
P_{0,2 x}(\l)-\delta_{ 2x, 0}&
=
\l\big(\beta\,F_{0,2 x-1}(\l)-(1-\beta)\,F_{0, 2 x+1}(\l)\big)
\,
\label{eq:Dirac1dRecursion2}.
\end{align}
It is easier to deal with a two-step recursion as it decouples the edges and vertex.
The recursion now stands as, for any $x$, 
\begin{align}
\label{eq:Recursionin1D}
p_{2 \t+2}(0,2x)& 
=
-\alpha \beta \,p_{2 \t}(0,2x-2)
-(1-\alpha ) (1-\beta )\,p_{2 \t}(0,2x+2) 
\nnn 
&
+\big( \a(1-\b)+\b(1-\a)\big)\,p_{2 \t}(0,2x)
\,, \\
f_{2 \t+1}(0,2x+1)& 
=
-\alpha \beta \,f_{2 \t-1}(0,2x-1)
-(1-\alpha ) (1-\beta )\,f_{2 \t-1}(0,2x+3)
\nnn
&
+\big( \a(1-\b)+\b(1-\a)\big)\,f_{2 \t-1}(0,2x+1)
\,,
\end{align}
and at the level of generating functions, we have
\begin{gather}
\label{eq:twosteprecursionlinevertex}
P_{0,2 x}(\l)- \d_{2x,0}
=
\lambda ^2 
\Big(
-\alpha  \beta  \,P_{0,2 x-2}(\l)
+ \big( \a(1-\b)+\b(1-\a)\big) P_{0,2 x}(\l)
-(1-\alpha ) (1-\beta )\,P_{0,2 x+2}(\l)
\Big)
\,,\\
F_{0,2 x+1}(\l)
=
\lambda ^2 \Big(
-\alpha  \beta  \,F_{0,2 x-1}(\l)
+
\big( \a(1-\b)+\b(1-\a)\big) F_{0,2 x+1}(\l)
-
(1-\alpha ) (1-\beta )\,F_{0,2 x+3}(\l)
\Big)
\,.
\end{gather}
Similarly to our analysis of the simple random walk, by carefully paying attention to the fact that the origin is special in the recursion relations, we introduce index forward and backward shift operators, $R$ and $Q$, for some $x$ such that
\al{
R P_{0,2x}(\l)&= P_{0,2x+2}(\l)
\,,
\iif x \geq 0 \, \label{eq:upshift2v} , \\
Q P_{0,2x}(\l)&= P_{0,2x-2}(\l)
\,,
\iif x \leq 0\label{eq:downshift2v}   
\,.
}
Solving for $P$, we will then be able to retrieve $F$ from Eq.~\eqref{eq:Dirac1dRecursion2}. On the positive half-line, we have
\be 
\label{eq:positiveCharacteristic}
\Bigg(
R  
- 
\lambda ^2
\Big(
-(1-\a) (1-\b ) R^2
+ 
\big(
\a(1-\b)+\b(1-\a)
\big) 
R 
-\a \b  
\Big) 
\Bigg)
P_{0,2x}(\l)
=0
\,,
\iif x \ge 0
\,,
\ee 
which gives us the characteristic equation
that has two solutions 
\be 
\label{eq:rPpm}
r_{P}{}_{\pm}(\l)=\frac{\lambda ^2 (-2 \a \b +\a+\b )\pm\sqrt{\lambda ^4 (\a-\b )^2+2 \lambda ^2 (\a (2 \b -1)-\b )+1}-1}{2 (\a-1) (\b -1) \lambda ^2}\,.
\ee 
We keep the solution regular at $\lambda=0$, rewriting it as $r_{P}{}_{+}(\l)=r_{P}(\l)$.
The recursion is solved by
\be
P_{0,2x}(\l)
=
r_{P}(\l)
^{x}
P_{0,0}(\l)
\, ,
\iif
x \ge 0\,.
\label{eq:positivepowerlaw}
\ee
Similarly, we can compute the characteristic roots for the negative half-line, $x\leq 0$. 
We obtain,
\be 
{\label{eq:negativeCharacteristic}
\Bigg(
Q 
- 
\lambda ^2
\Big(
-\a \b  Q^2
+
\big(
\a(1-\b)+\b(1-\a)
\big) 
Q
-(1-\a) (1-\b )
\Big) 
\Bigg)
P_{0,2x}(\l)
=
0
\,,}
\ee
with solutions 
\be 
q_{P\pm}(\l) = \frac{\lambda ^2 (-2 \a \b +\a+\b )-1 \pm \sqrt{\lambda ^4 (\a-\b )^2+2 \lambda ^2 (\a (2 \b -1)-\b )+1}}{2 \a \b  \lambda ^2}\, .
\ee
If we pick up the solution $q_{P +}(\l)$ which is regular for $\l \to 0$, we obtain by setting $q_{P +}(\l) =q_{P}(\l) $, 
\be 
P_{0,2x}(\l) = q_{P}(\l)^{|x|}P_{0,0}(\l) 
\,,
\quad \text{if}\, x \leq 0
\,.
\label{eq:negativepowerlaw}
\ee
We solve for $P_{0,0}(\l)$ by substituting $P_{0,\pm 2}(\l)$ in terms of $P_{0,0}(\l)$ using \eqref{eq:positivepowerlaw} and \eqref{eq:negativepowerlaw} in \eqref{eq:twosteprecursionlinevertex}
\al{
P_{0,0}(\l)
&=
1
+
\lambda ^2 P_{0,0}(\l)
\Big(
-
\alpha \beta q_{P}(\l)
-
(1-\alpha ) (1-\beta ) r_{P}(\l)
+
\big(\alpha  (1-\beta )+(1-\alpha ) \beta \big)
\Big) 
\nnn
&= 
\frac{
1
}{
\sqrt{
\lambda ^4 (\alpha -\beta )^2+2 \lambda ^2 (\alpha  (2 \beta -1)-\beta )
+
1}
}\, .
}
We can then compute the vertex to edge transition generating function using \eqref{eq:Dirac1dRecursion1}. 
Ultimately, we are interested in computing powers of the Dirac operator, which count the sum of unbiased signed walks. Therefore, we will be interested when $\a = \b = \frac{1}{2}$. For this case, the characteristic equation induced by $R$ and $Q$ are equivalent. We obtain
 \begin{gather}
     q_{P}(\l) = r_{P}(\l) = \frac{\lambda ^2+2 \sqrt{1-\lambda ^2}-2}{\lambda ^2}
     \,,
     \\
     P_{0,0}(\lambda)
    =
    \frac{1}{\sqrt{1-\lambda^2}}\,.
 \end{gather}
All in all the generating functions are with $\alpha = \beta = 1/2$,
\al{ 
\label{eq:PxVtoVline}
P_{0,2x}(\l) 
&= q_P(\l)^{|x|} P_{0,0}(\l) 
= \frac{
\left(
\frac{
\lambda ^2
+
2 \sqrt{
1-\lambda ^2
}
-2
}
{
\lambda ^2
}
\right)
^{|x|}
}
{
\sqrt{
1-\lambda ^2
}
}
\,,
\quad \forall x
\, , 
\\
\label{eq:FxVtoEline}
F_{0,\pm2x\pm1}(\l)
&= 
\pm
\frac{
\left(
\sqrt{
1-\lambda ^2
}
-1
\right) 
\left(
\frac{
\lambda ^2
+2
\sqrt{
1-\lambda ^2
}
-2
}
{
\lambda ^2
}
\right)
^{x}
}
{
\lambda  \sqrt{
1-\lambda ^2
}
} 
\,,
\quad \text{if}
\,\, x \geq 0 
\,.
}


\subsubsection{Dirac walk on the $d \ge 3$ Bethe lattice}

\noindent
{\bf \underline{Dirac walks starting from a vertex on the Bethe lattice.}}
\vskip 5pt

\begin{figure}[H]
\centering
\begin{minipage}[t]{.75\textwidth}
\centering
\includegraphics[trim = 0.1cm 1.1cm 0.1cm 0.1cm,clip,width=8.5cm]{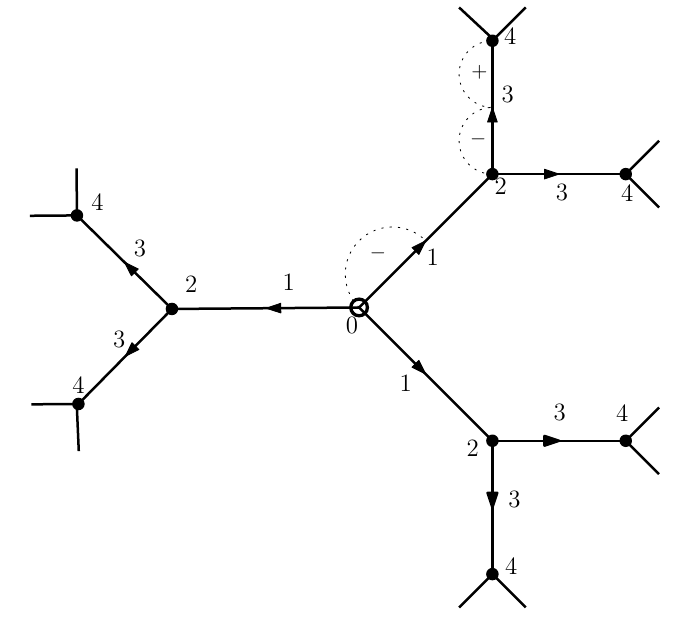}
\caption{A $d=3$ Bethe lattice with sites labeled in units of half-edge length away from the origin,
circled and denoted with $0$.
We show the directions assigned on edges in arrows.
}
\label{fig:directedbethe3reg}
\end{minipage}
\hfill
\centering
\begin{minipage}[b]{.75\textwidth}
\centering
\includegraphics[width=7.5cm]{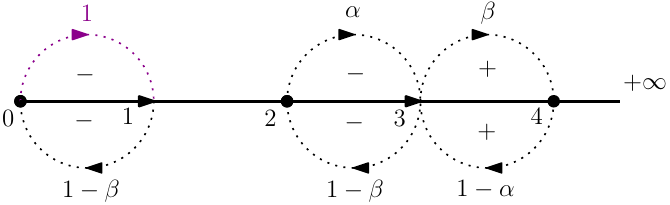}
\caption{
The half-line which the Bethe lattice (e.g., Figure~\ref{fig:directedbethe3reg}) is mapped onto.
We show the transition weights for some steps and the (anomalous) weight $-1$ on the step between vertices $0$ and $1$ reflects the boundary condition \eqref{eq:RecursionBCposstepBetheEdge}.
}
\label{fig:halflinemapped}
\end{minipage}
\label{fig:drectedBethe}
\end{figure}

\vskip 5pt
\noindent
To study the Dirac walk, we need to assign orientations to the edges of the Bethe lattice. 
We choose a root vertex, such that all the edges are pointing away from the root vertex. 
For the discussion of a walker starting from a vertex we choose the root to be the starting point of our random walker and call it the origin. 
Unlike the case of the simple random walk, where distance was measured in terms of units of one edge length, in the case of the Dirac walk, we measure distance in units of half-edge length. 
We observe that we have $d(d-1)^{x-1}$ vertices at distance (i.e. the number of half-edges) $2x$ away from the origin, with $x\in \mathbb{Z}_{\geq 0}$. 
Like our discussion in Section \ref{RandomwalkBethe}, we map a vertex or an edge at  distance $x$ away from the origin of the Bethe lattice to an edge or a vertex distance $x$ away from the origin of the half-line. 
Similarly, we have $d(d-1)^{x}$ edges,  distance $(2x+1)$ away from the origin of the Bethe lattice.
For $x \neq 0$, if we are on a vertex $2x$ distance from the origin of the Bethe lattice, there are $d-1$ ways to go further away from the origin in a step and one way to go closer to the origin. But if we are on an edge, there is only one way to go away and one way to come closer to the origin. This is pictorially represented in Figure \ref{fig:directedbethe3reg}.

Thus, our problem of computing transition weights can be mapped to a biased Dirac walker on the half-line with $\a = \frac{d-1}{d}$ and $\b = \frac{1}{2}$, and for $x=0$ the walker ends up at edge $1$ with weight $-1$. We will present the result for a general $\a$ and $\b$ and plug in the suitable values later. Starting from $0$,  we get the recursion
\begin{align}
\label{eq:RecursioninstepandposBethe}
f_{2\t+1}(0,2 x+1)&=(1-\a)  p_{2 \tau }(0,2 x+2)-\a  p_{2 \tau}(0,2 x)\,, \quad \text{if } x\geq 1 \,, \\
p_{2 \tau +2}(0,2x)&=\b f_{2 \tau +1}(0,2 x-1)-(1-\b)  f_{2 \tau +1}(0,2 x+1)\,,\quad \text{if } x\geq1\,.
\end{align}
This needs to be supplemented with reflecting boundary conditions representing a walker starting from $0$ going to edge $1$ with weight $-1$
\begin{align}
\label{eq:RecursionBCposstepBethe}
p_{2 \tau +2}(0,0)&=-(1-\b)  f_{2 \tau +1}(0,1)\,,  \\
\label{eq:RecursionBCposstepBetheEdge}
f_{2 \tau +1}(0,1)&=(1-\a) p_{2 \tau}(0,2)-p_{2 \tau}(0,0)\,,\\
p_{0}(0,2x) &= \delta_{0,2x}\,. 
\end{align}
At the level of generating functions, we obtain
\begin{align}
F_{0,2x+1}(\l)&=\l\left((1-\a) P_{0,2 x+2}(\l)-\a  P_{0,2 x}(\l)\right)\,, \quad \text{if }  x\geq 1\, \label{eq:f2xplus1}\\
P_{0,2 x}(\l)
&=
\l
\left(
\b F_{0,2 x-1}(\l)
-
(1-\b)F_{0,2 x+1}(\l)
\right)
\,,
\quad \text{if } x \geq 1 \,,
\\
P_{0,0}(\l)-1&=-\l\, (1-\b) F_{0,1}(\l)\,, \\
F_{0,1}(\l)&=\l\left((1-\a) P_{0,2}(\l)-P_{0,0}(\l)\right)\,.
\end{align}
As in Section \ref{ssec:DiracWalkonline}, to decouple the odd and even steps, we will solve the recursion in two steps for $P$, for $x \geq 2$,
\be 
\label{eq:twosteprecursionBulk}
P_{0,2 x}(\l)
=
\lambda ^2 
\Big(
-\alpha  \beta  \,P_{0,2 x-2}(\l)
+ \big( \a(1-\b)+\b(1-\a)\big) P_{0,2 x}(\l)
-(1-\alpha ) (1-\beta )\,P_{0,2 x+2}(\l)
\Big)
\,,
\ee
and retrieve $F$ from eq.~\eqref{eq:f2xplus1}. Notice that the recursion is the same as eq.~\eqref{eq:twosteprecursionlinevertex}, 
but the boundary conditions differ
\begin{align}
\label{eq:twostepBC1}
P_{0,0}(\l)-1
&=
\lambda ^2 \Big(
(1-\b) P_{0,0}(\l)
-(1-\a)(1-\b) P_{0,2}(\l)\Big) 
\,, 
\\
\label{eq:twostepBC2}
P_{0,2}(\l)
&=
\lambda ^2 \Big(
-\b  P_{0,0}(\l)
+\big(\a(1-\b)+\b(1-\a)\big)P_{0,2}(\l)
-(1-\a)(1-\b) P_{0,4}(\l)
\Big)
\,.    
\end{align}
since the magnitude of the weight from the site $0$ to the site $1$ is $1$ rather than $\alpha$. Hence, we obtain again the form
\be 
P_{0,2x}(\l) = r_{P}(\l)^{x-1}P_{0,2}(\l) 
\,,
\quad \text{if } x \geq 1
\,.
\ee
with $r_P(\l)$ given by $r_{P+}$ in eq.~\eqref{eq:rPpm}.
For a general $\a$ and $\b$ we can solve for $P_{0,0}(\l)$ and $P_{0,2}(\l)$ using the boundary conditions.
\be
\label{eq:P00}
P_{0,0}(\l)-1=\lambda ^2 ((1-\beta ) P_{0,0}(\l)-(1-\alpha ) (1-\beta ) P_{0,2}(\l))
\,,
\ee
and
\be
P_{0,2}(\l) = \l^2\left( -\beta P_{0,0}(\l) + (\a(1-\b)+\b(1-\a)) P_{0,2}(\l)  -(1-\a)(1-\b)P_{0,2}(\l)r_{P}(\l)\right) .
\ee
We solve for $P_{0,2}(\l)$ as a function of $P_{0,0}(\l)$,
\be
\label{eq:P02}
P_{0,2}(\l) 
= 
\frac
{-\beta  \, \lambda ^2 P_{0,0}(\l)}
{
1
+ 
\lambda^2
\big(
2  \alpha   \beta  
-\alpha  
-\beta  
+ (\alpha \, \beta 
-\alpha 
-\beta  
+ 1) r_{P}(\lambda )
\big)
}
\,.
\ee
Plugging in $P_{0,2}(\l)$ in \eqref{eq:P00}, we get
\begin{align}
P_{0,0}(\l) 
=
\frac{
\lambda ^2 (\alpha -\beta )
+
\sqrt{
\lambda ^4 (\alpha -\beta )^2
+
2 \lambda ^2 (\alpha  (2 \beta -1)-\beta )
+
1
}
-
2 \alpha 
+
1
}
{
2 (1-\alpha)
\left(
1
-
\lambda ^2
\right)}
\,.
\end{align}
Replacing $\a = \f{d-1}{d}$ and $\b = \f{1}{2}$, we get
\begin{equation}
\label{eq:PDiracBL}
    P_{0,2x}(\l)= \begin{dcases}\frac{4 (d-1)}{2 \left(\lambda ^2-2\right)+d \left(\sqrt{\frac{(d-2)^2 \lambda ^4}{d^2}-4 \lambda ^2+4}-\lambda ^2+2\right)}\,, & \text{if } x=0\,,\\
    \frac{4 d \left(\frac{d \left(\sqrt{\frac{(d-2)^2 \lambda ^4}{d^2}-4 \lambda ^2+4}+\lambda ^2-2\right)}{2 \lambda ^2}\right)^x}{2 \left(\lambda ^2-2\right)+ d \left(\sqrt{\frac{(d-2)^2 \lambda ^4}{d^2}-4 \lambda ^2+4}-\lambda ^2+2\right)}\,, & \text{if } x\geq 1\,,
    \end{dcases}
\end{equation}
and from \eqref{eq:f2xplus1} it follows
\be 
F_{0,2x+1}(\l) 
= \frac{4 d \lambda  \left(\frac{d \left(\sqrt{\frac{(d-2)^2 \lambda ^4}{d^2}-4 \lambda ^2+4}+\lambda ^2-2\right)}{2 \lambda ^2}\right)^x}
{2 \lambda ^2 - d \left(\sqrt{\frac{(d-2)^2 \lambda ^4}{d^2}-4 \lambda ^2+4}-\lambda ^2+2\right)} 
\,,
\quad
\text{if } x\geq 0
\label{eq:PxdiracbetheVtoE}
\,.
\ee

As in the case of the simple random walk, generating functions \eqref{eq:PDiracBL} and \eqref{eq:PxdiracbetheVtoE} contain information about the transition weight to reach any one of the sites at certain distance. We want the information about the transition weight of reaching a particular site on a Bethe lattice. That can be obtained by the generating functions $\hat{P}_{0,2x}(\l) = \frac{P_{0,2x}(\l)}{d(d-1)^{x-1}}$ and $\hat{F}_{0,2x+1}(\l) = \frac{F_{0,2x+1}(\l)}{d(d-1)^{x}} $. Explicitly, they are
\begin{gather}
\label{eq:VertexStartingDiracBethe}
\hat{P}_{0,2x}(\l)
=\frac{4 (d-1) \left(\frac{d \left(\sqrt{\frac{(d-2)^2 \lambda ^4}{d^2}-4 \lambda ^2+4}+\lambda ^2-2\right)}{2 (d-1) \lambda ^2}\right)^x}{\left(d \left(\sqrt{\frac{(d-2)^2 \lambda ^4}{d^2}-4 \lambda ^2+4}-\lambda ^2+2\right)+2 \left(\lambda ^2-2\right)\right)}\,,
 \quad \text{if }  x \ge 0
\,, 
\\
\label{eq:VertexStartingDiracBetheF}
\hat{F}_{0,2x+1}(\l)
=\frac{4 \lambda  \left(\frac{d \left(\sqrt{\frac{(d-2)^2 \lambda ^4}{d^2}-4 \lambda ^2+4}+\lambda ^2-2\right)}{2 (d-1) \lambda ^2}\right)^x}
{2 \lambda ^2 - d \left(\sqrt{\frac{(d-2)^2 \lambda ^4}{d^2}-4 \lambda ^2+4}-\lambda^2 +2\right)}
\,, \quad
\text{if } x\geq 0
\,.
\end{gather}

\noindent
{\bf \underline{Dirac walks starting from an edge on the Bethe lattice.}}

\vskip 5pt
\noindent
We also want to compute the generating functions for the transition weight of a Dirac walker starting from a particular edge to another site.
Let's choose a fiducial edge attached to the root, call it the origin and label it $0$. 
Our aim is to map the Bethe lattice to a one-dimensional system, similarly to our previous calculations.

\begin{figure}[H]
\centering
\begin{minipage}[t]{.75\textwidth}
\centering
\includegraphics[trim=0.2cm 1.1cm 0.2cm 1.1cm,clip,width=9cm]{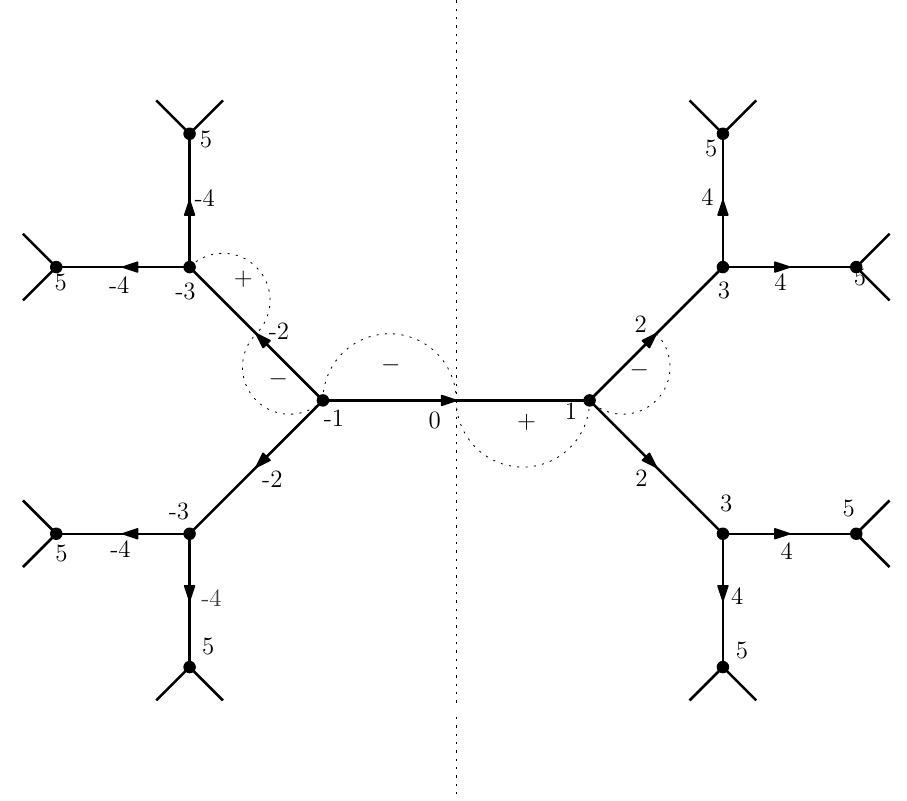}
\caption{A $d=3$ Bethe lattice with an edge attached to the root vertex labeled $0$ as the origin of the walk and sites labeled in units of half-edge length directed away from the origin. The two steps tracing $0$, $-1$ and $-2$ have consecutive $-$ signs, while for any other two neighboring sites e.g., steps tracing $-1$, $-2$ and $-3$; and $0$, $1$ and $2$ have alternate $+$ and $-$ signs.
}
\label{fig:edgestart}
\end{minipage}
\hfill
\centering
\begin{minipage}[b]{.75\textwidth}
\centering
\includegraphics[width=10.0cm]{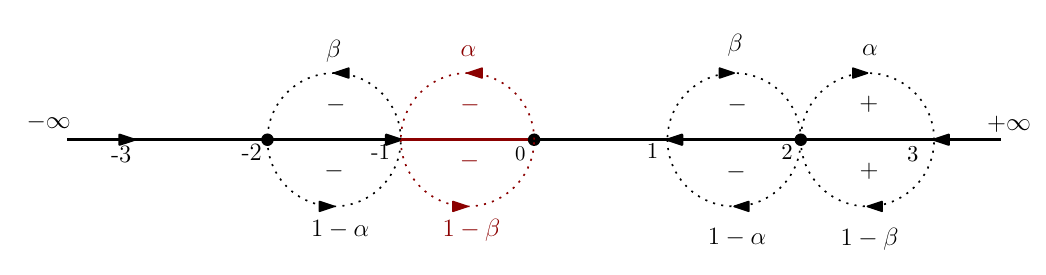}
\caption{
A  directed line with associated weights for steps between edges and vertices on which Figure \ref{fig:edgestart} has been mapped to. The anomalous half of the edge labeled $-1$ has been colored dark red and has been assigned a $-$ sign. This ensures that the weight of steps tracing $0$, $-1$ and $-2$ has a $+$ sign, matching with the Bethe lattice.}
\label{fig:fulline}
\end{minipage}
\end{figure}

The previous computation, when the walk started at the root vertex of the Bethe lattice, benefited from a symmetry that allowed to map the random process to a walk on the half-line. A walk starting from any other point will a priori break this symmetry.
We resolve this issue by mapping the Bethe lattice to the integer line $\mathbb{Z}$, Figure~\ref{fig:fulline} and Figure \ref{fig:edgestart}. 
The edge $0$ divides the Bethe lattice into two halves, the left one for which the first step away from the origin has a $``-"$ sign and the right one for which the first step away from the origin has a $``+"$ sign.

On the Bethe lattice (cf. Figure \ref{fig:edgestart}), a walker starting from an edge, different from the origin edge, has one way to go further away from the origin, out of the two steps it can take. From a vertex, it has $d-1$ ways to go further away from the origin out of $d$ possible steps. As earlier, such process can be mapped to a walker on the line with a drift away from the origin in both directions. 
Except for steps between the sites $-1$ and $0$, the weight of a step away from the origin to the nearest edge is $\a = \f{1}{2}$ and $-(1-\a) = -\f{1}{2}$ towards the origin. From an edge, a step away from the origin will bare weight $-\b = -\f{d-1}{d}$ and $(1-\b) = \f{1}{d}$ towards the origin.
This is illustrated with a few examples in Fig.~\ref{fig:edgestart}. 
Let us first notice the orientation defect with respect to a walk starting at the root vertex.
If we go from $0$ to a vertex $-2$ in two steps, the Dirac walker has a $``+"$ sign. Meanwhile, any other two consecutive steps away from 0 will contribute a $``-"$ sign. This defect has to be reflected in special conditions for the recursion between sites $-1$ and $0$.

We define $p^{(E)}_{2\t}(0,2x)$ as the weight of a Dirac walker on the line, that starts from the origin $0$ of the line and ends on a vertex $2x$ on the line. This corresponds to the transition weight of a Dirac walker on a Bethe lattice, starting from the edge $0$ and ending on any of the $(d-1)^{x}$ edges with label $2x$. The superscript $(E)$ serves to emphasize that the origin of the Bethe lattice is an edge.

The corresponding generating function and two-step recursion (for $x\geq 2$) are
\begin{gather}
P^{(E)}_{0,2x}(\l)= \sum^{\infty}_{\t=0}\l^{2\t}p^{(E)}_{2\t}(0,2x) \quad \text{if } x \in \mathbb{Z}\,,\\
P^{(E)}_{0,\pm 2 x}(\l)
=
\lambda ^2 
\Big(
-\alpha  \beta  \,P^{(E)}_{0,\pm(2 x-2)}(\l)
+\big( 
\a(1-\b)+\b(1-\a)
\big) 
P^{(E)}_{0,\pm 2 x}(\l)
-
(1-\alpha ) (1-\beta )\,P^{(E)}_{0,\pm(2 x+2)}(\l)
\Big)
\,, 
\label{eq:recedgebulk}
\end{gather}
supplemented with equations for the generating functions at the sites $\pm 2$:
\be 
\label{eq:rec-edges}
P^{(E)}_{0,\mp 2}(\lambda) 
= 
\l^2\Big(
\pm \alpha\beta P^{(E)}_{0,0}(\lambda) 
+ \big(\a(1-\b)+\b(1-\a)\big)P^{(E)}_{0,\mp 2}(\lambda) 
-(1-\alpha)(1-\beta) P^{(E)}_{0,\mp 4}(\l)
\Big)
\,,
\ee
where the coefficient of $P^{(E)}_{0,0}(\l)$ is $+\a\b$ for the recursion of $P^{(E)}_{0,-2}(\l)$ to take into account that any two steps from $-2$ to $0$ will have a positive sign differing from $2$ to $0$;
and at the site $0$, we have:
\be
\label{eq:rec-edges0}
P^{(E)}_{0,0}(\lambda) -1
= 
\l^2\Big(
(1-\alpha)(1-\beta)P^{(E)}_{0,-2}(\lambda) 
+ 
2\a(1-\b)P^{(E)}_{0,0}(\lambda) 
-
(1-\alpha)(1-\beta)P^{(E)}_{0,2}(\lambda)
\Big)
\,.
\ee
Let us detail the coefficients. The ways in which a walker residing at $0$ can return to $0$ in two steps are that they can go back and forth to $\pm 1$. For going to $-1$ the walker has the weight of $-\a$ and it has one out of $d-1$ ways to return back to $0$. Thus it can move back from $-1$ to $0$ with transition weight $-(1-\b)$. For going to $1$ and returning back to $0$, we have weight $\a$ to go to $1$ and weight $(1-\b)$ to return to $0$ from $1$. Hence, the coefficient of $P^{(E)}_{0,0}(\l)$ is $2\a(1-\b)$.
We define the index shift operator $R$ such that
\be 
R P_{0,\pm 2x}(\l) 
=
P_{0,\pm(2x+2)}(\l)
\, ,
\iif 
x \ge 1
\, ,
\ee
which allows us to write the recursion~\eqref{eq:recedgebulk} for 
$x \ge 1$ as
\be
\label{eq:characteristicRootBetheedgeplus}
\Bigg(
R 
- 
\l^2
\Big(
- 
(1-\a)(1-\b) R^2 
+ 
\big(
\a(1-\b)+\b(1-\a)
\big)
R
-\a\b
\Big)
\Bigg)
P^{(E)}_{0,\pm 2x}(\l)=0
\,,
\ee
which leads to the same characteristic equations as for the vertex to vertex Dirac walk Eq.~\eqref{eq:positiveCharacteristic} 
implying
\be
P^{(E)}_{0, \pm 2x}(\l) = (r_{P}(\l))^{x-1}P^{(E)}_{0,\pm 2}(\l)
\,,
\iif 
x \ge 1\,,
\ee
for $r_P(\l)$ given in Eq.~\eqref{eq:rPpm}.
Now, adding the two equations in Eq.~\eqref{eq:rec-edges},
it follows that $P^{(E)}_{0,-2}(\lambda) = -P^{(E)}_{0,2}(\lambda)$. 
This relationship allows us to solve for $P^{(E)}_{0,\pm 2}(\lambda)$ and $P^{(E)}_{0,0}(\lambda)$ from \eqref{eq:rec-edges0}, resulting in:
\be
P^{(E)}_{0,0}(\lambda) = \frac{1}{\lambda^2 (\beta -\alpha )+\sqrt{\lambda^4 (\alpha -\beta )^2+2 \lambda^2 (\alpha (2 \beta -1)-\beta )+1}}. \nonumber
\ee
For $\alpha = \frac{1}{2}$ and $\beta = \frac{d-1}{d}$, we obtain:
\begin{align}
\label{eq:EdgetoEdge}
P^{(E)}_{0,\pm 2x}(\l) 
&= \pm 
\frac{2 d 
\left(\frac{d \left(\sqrt{\frac{(d-2)^2 \lambda^4}{d^2}-4 \lambda^2+4}+\lambda^2-2\right)}{2 \lambda^2}\right)^{x}}
{d \sqrt{\frac{(d-2)^2 \lambda^4}{d^2}-4 \lambda^2+4}+(d-2) \lambda^2}
\,,
\quad
 {\rm if} \;  x \ge 1
\,.
\end{align}
The generating function for reaching a particular edge on the Bethe lattice is \footnote{As a caution, let us emphasize that the indices on $P^{(E)}$ refer to vertices on the line, whereas those on $\hat{P}^{(E)}$ refer to edges on the Bethe lattice.}
\begin{align}
\label{eq:EdgetoEdgehat}
\hat{P}^{(E)}_{0,0}(\l) 
&=  
\frac{2 d}
{d \sqrt{\frac{(d-2)^2 \lambda^4}{d^2}-4 \lambda^2+4}+(d-2) \lambda^2}
\,,\\
\hat{P}^{(E)}_{0,\pm 2x}(\l) 
&= \f{P^{(E)}_{0,\pm 2x}(\l)}{(d-1)^{x}}
= \pm 
\frac{2 d 
\left(\frac{d \left(\sqrt{\frac{(d-2)^2 \lambda^4}{d^2}-4 \lambda^2+4}+\lambda^2-2\right)}{2 (d-1) \lambda^2}\right)^{x}}
{d \sqrt{\frac{(d-2)^2 \lambda^4}{d^2}-4 \lambda^2+4}+(d-2) \lambda^2}
\,,
\quad
{\rm if} \;  x \ge 1
\,.
\end{align}

We define the weight of starting from $0$ and arriving at an edge $2x+1$ on the line (corresponding to vertices on the Bethe lattice) to be $f^{(E)}_{2\t+1}(0,2x+1)$, with the corresponding generating function $F^{(E)}_{0,2x+1}(\l)$.
The one step recursions in odd number of steps are given by 
\begin{gather}
\label{eq:EdgetoVertex}
F^{(E)}_{0,2x+1}(\l) 
= \l
(
\a\,P^{(E)}_{0,2x}(\l)
- 
(1-\a)\,P^{(E)}_{0,2x+2}(\l)
) 
\,,
\iif x \ge 0 
\, ,\\
F^{(E)}_{0,-(2x+1)}(\l) 
= \l
(
\a\,P^{(E)}_{0,-2x}(\l)
- 
(1-\a)\,P^{(E)}_{0,-(2x+2)}(\l)
) 
\,,
\iif x \ge 1 
\, ,
\end{gather}
and the defective oriented edge implies
\be
F^{(E)}_{0,-1}(\l) 
= \l
(
-\a\,P^{(E)}_{0,0}(\l)
- 
(1-\a)\,P^{(E)}_{0,-2}(\l)
) 
\,.
\ee
We can use \eqref{eq:EdgetoVertex} and \eqref{eq:EdgetoEdge} and compute
\be
F^{(E)}_{0,\pm(2x+1)}(\l) 
=\mp \f{F_{0,2x+1}(\l)}{2}
\,,
\quad
{\rm if}\,\,x \ge 0
\, .
\ee
The generating function for reaching a particular vertex on the Bethe lattice is
\be
\label{eq:EdgetoVertexhat}
\hat{F}^{(E)}_{0,\pm (2x+1)}(\l)
=\frac{F^{(E)}_{0,\pm(2x+1)}(\l) }{(d-1)^{x}}
=\mp \f{d}{2}\hat{F}_{0,2x+1}(\l)\,,
\quad
{\rm if}\,\,x \ge 0 
\, .
\ee

\subsection{Shift of starting point}

The computation of transition probabilities of the simple random walk and transition weights of the Dirac walks on the line $\mathbb{Z}$ has been performed with the starting point coinciding with the origin. We would like to generalize the computation to any starting point. 

For a simple random walker, the homogeneity of the Bethe lattice implies that between any two vertices $x$ and $y$, the transition probabilities depend only on the distance $g(x,y)$\footnote{In this subsection, we will use the abuse of the notation $\omega_{\t,0\to\abs{g(x,y)}}$ to signify a walk from the origin to a site at distance $\abs{g(x,y)}$, for sites $x$ and $y$.}
\be
\hat{h}_{\t}(x,y) = \begin{dcases}
     h_{\t}(0,|g(x,y)|) & \text{if } d= 2\,,\\
     \hat{h}_{\t}(0,\abs{g(x,y)}) &\text{otherwise.}
\end{dcases}
\ee 

Let us discuss some general facts about Dirac walkers on $d$-regular trees.
First, the number of walks with a given number of steps is independent of the location of the starting point but depends on whether it is an edge or a vertex.
Between two sites $z$ and $w$, any Dirac walk $\omega_{\t,z\to w}$ consists of $|g(z,w)|$ steps tracing the geodesic $g(z,w)$, and the remaining $\t - |g(z,w)|$ steps are distributed as closed walks on each of the lattice sites on the geodesic. For Dirac walks $\omega_{\t,0\to |g(z,w)|}$, we similarly have at least $|g(z,w)|$ steps tracing the $g(0,|g(z,w)|)$ \footnote{We used the fact that $|g(z,w)|= |g(0,|g(z,w)|)|$.}, and the rest $\t - |g(z,w)|$ steps are distributed as closed walks on each of the lattice sites on the geodesic $g(0,|g(z,w)|)$. Thus,
\be
|\Omega_{\t,z\to w}| = |\Omega_{\t,0\to |g(z,w)|}|. 
\ee
Second, closed walks have positive sign, since on a tree, every half-edge has to be traversed in both directions to return at the origin and the sign of the step is independent of the direction of the step.
Thus, for every closed walk of an even number of steps $2 k$,
\be
\sgn(\omega_{2k,2x\to2x}) = 1\, .
\ee
Therefore, the overall sign of a particular Dirac walk depends only on the sign of the walk on the geodesic $g(z,w)$ connecting two sites.

Third, we can see that for each step from $2x \to 2x\pm 1$, \footnote{The notation $\pm$ signifies here that consecutive steps have been made in the same direction, either positive or negative.} the sign of the step has opposite sign compared to the step from $2x\pm 1 \to 2x\pm 2$, in other words $\sgn(\omega_{1,2x\to2x\pm 1})=-\sgn(\omega_{1,2x\pm 1\to2x\pm 2})$. This implies, 
\be
\sgn(\omega_{2,2x\to2x\pm 2}) = -1\, .
\ee
We then conclude
\be
\sgn(\omega_{2\t,2x\to2 y}) =
\begin{dcases}
    -1 &  \text{if}\, |g(2x,2y)| \equiv 2 \mod{4}\\
    1 & \text{if}\, |g(2x,2y)| \equiv 0 \mod{4} \, .
\end{dcases}
\ee 
For a Dirac walk from a vertex to a vertex, $\omega_{2\t,0\to|g(2x,2y)|}$, we also have
\be
\sgn(\omega_{2\t,0\to |g(2x,2y)|})=
\begin{dcases}
    -1 &  \text{if}\, |g(2x,2y)| \equiv 2 \mod{4}\\
    1 & \text{if}\, |g(2x,2y)| \equiv 0 \mod{4} \, .
\end{dcases}
\ee 
Since for any two sites $z$ and $w$, the number of walks does not depend on the origin, i.e., $|\Omega_{\t,z\to w}| = |\Omega_{\t,0\to |g(z,w)|}|$, it is enough to consider only geodesic walks in the following.

\subsubsection{Shift of starting point for the Dirac walk on the line}

\vskip 5pt
\noindent
{\bf {From vertex to vertex and edge to edge.}}
For the case of the Dirac walk on the line, every vertex has one incoming edge and one outgoing edge, as we see in Figure~\ref{fig:directedline}. Again, since two steps in any fixed direction starting from any lattice site will have a sign $-1$, the transition weight between two sites of the same nature only depends on their geodesic distance $2\gamma$, such that
\begin{gather}\forall x,y,z,w \in \mathbb{Z}, \gamma \in \mathbb{N}: |g(2x,2y)|= |g(2z+1,2w+1)|=2\gamma\,,\\
p_{2\t}(2x,2y) = f_{2\t}(2z+1,2w+1) = p_{2\t}(0,2\gamma)
\,.
\end{gather}

\vskip 5pt
\noindent
{\bf {From vertex to  edge and edge to vertex.}}
In this case as well, the number of walks only depends on the geodesic distance between the end points of the walk. When sites $2x, 2y+1, 2w$, and $2z+1$ obey $|g(2w,2z+1)|=|g(2x,2y+1)|=\d\in \mathbb{N}$, we would like to compare $p_{2\t+1}(2z+1,2w)$ and $f_{2\t+1}(2x,2y+1)$ to $f_{2\t+1}(0,\d)$, that is, to walks with an endpoint on the positive side of the line in Figure~\ref{fig:directedline}.
\begin{figure}[H]
    \centering
    \begin{minipage}[b]{.8\textwidth}
    \includegraphics[width=13cm]{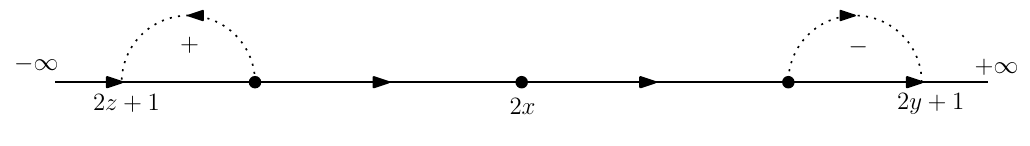}
    \caption{A directed line with an arbitrary starting vertex $2x$ for the Dirac walk. If the ending point is $2y+1$, such that $2x<2y+1$, the last step on the geodesic walk takes a $-$ sign. 
    On the other hand, if the ending point is  $2z+1$, where $2z+1 < 2 x$, the last step on the geodesic walk to the ending edge takes a $+$ sign. 
    }
    \label{fig:directedlineVtoE}
    \end{minipage}
\end{figure}
First, let us compare $f_{2\t+1}(2x,2y+1)$ and $f_{2\t+1}(0,\d)$. For the walk from $0$ to $\d$, the last step is towards $+\infty$ and has sign $-1$.
Moreover, the sign of a walk from $2x$ to $2y+1$ would be determined by the distance between the underlying vertex to vertex walk from $2x$ to the nearest vertex to $2y+1$ on the geodesic $g(2x,2y+1)$ and the sign of the last step on $g(2x,2y+1)$. The latter sign depends on the relative positions of $2x$ and $2y+1$. If $2x < 2y+1$, the last step is towards $+\infty$, hence has a sign $-1$, otherwise it is towards $-\infty$, bringing a sign $+1$ (cf. Figure~\ref{fig:directedlineVtoE}). 
Therefore, the signs relate by 
\be
\sgn(\omega_{2\t+1,2x\to 2 y+1}) =
\begin{dcases}
    \sgn(\omega_{2\t+1,0\to \d}) &\text{if } 2x < 2y+1\,,\\
    -\sgn(\omega_{2\t+1,0\to \d}) &\text{otherwise,}
\end{dcases}
\ee 
and the transition weights by
\be
f_{2\t+1}(2x,2y+1) =
\begin{dcases}
    f_{2\t+1}(0,\d) &\text{if } 2x < 2y+1 \,,\\
   -f_{2\t+1}(0,\d) &\text{otherwise.}
\end{dcases}
\ee

\begin{figure}[H]
    \centering
    \begin{minipage}[b]{.8\textwidth}
    \includegraphics[width=13cm]{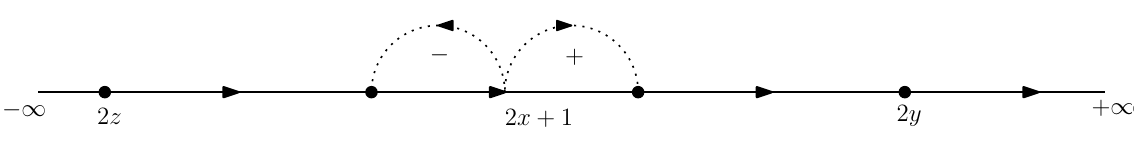}
    \caption{
    A directed line with an arbitrary starting edge $2x+1$ for the Dirac walker. If the ending point is $2y$, where $2x+1 < 2y$, the first step from the starting edge on the geodesic walk is towards $+\infty$, hence it takes a $+$ sign. Otherwise, if the ending point is $2z$, the first step on the geodesic walk from the starting edge takes a $-$ sign. 
    }
    \label{fig:directedlineEtoV}
    \end{minipage}
\end{figure}

Second, for the walk between $2z+1$ and $2w$, the sign of the walk would be determined by the product of the sign of the first step and the sign of the remaining vertex to vertex walk. If $2z+1< 2w$, the first step is towards $+\infty$, hence a sign $+1$, otherwise, it is towards $-\infty$, hence a sign $-1$, as shown in Figure~\ref{fig:directedlineEtoV}. 
Therefore, the signs relate as follows
\be
\sgn(\omega_{2\t+1,2z+1\to 2w}) =
\begin{dcases}
    -\sgn(\omega_{2\t+1,0\to\d}) &\text{if } 2z+1< 2w \,,\\
    \sgn(\omega_{2\t+1,0\to \d}) &\text{otherwise,}
\end{dcases}
\ee 
and the transition weights are
\be
p_{2\t+1}(2z+1,2w) =
\begin{dcases}
    -f_{2\t+1}(0,\d) &\text{if } 2z+1< 2w \,,\\
   f_{2\t+1}(0,\d) &\text{otherwise.}
\end{dcases}
\ee

\noindent
{\bf {Summary.}}

\vskip 5pt
\noindent
As the arguments stand for general $\t$, for all $ x,y,z,w \in \mathbb{Z}$ and $\gamma \in \mathbb{N}$, such that the distance between the pairs is $|g(2x,2y)|= |g(2z+1,2w+1)|=2\gamma$, the generating functions obey the relations
\begin{gather}
\label{eq:SRWlinegeneral}
H_{x,y}(\l) = H_{0,|g(x,y)|}(\l)
\,,
\\
\label{eq:VtoVEtoEDWgeneral}
P_{2x,2y}(\l) = F_{2z+1,2w+1}(\l) = P_{0,2\gamma}(\l) 
\, ,
\\
\label{eq:VtoElinegeneral}
F_{2x,2y+1}(\l) =
\begin{dcases}
    F_{0,|g(2x,2y+1)|}(\l) &\text{if } 2x \in g(0,2y+1) \,,
    \\
   -F_{0,|g(2x,2y+1)|}(\l) &\text{otherwise,}
\end{dcases}
\\
\label{eq:EtoVGeneral}
P_{2y+1,2x}(\l)=
\begin{dcases}
    -F_{0,|g(2x,2y+1)|}(\l) &\text{if }  2y+1 \in g(0,2x) \,,\\
   F_{0,|g(2x,2y+1)|}(\l) &\text{otherwise.}
\end{dcases}
\end{gather}

\subsubsection{Shift of starting point for the Dirac walk on the $d\geq 3$ Bethe lattice.}
\label{sec:originshift}

Regarding degree $d\geq 3$, let us take a look at Figure~\ref{fig:directedbethe3reg}. We see that with our choice of edge orientations of the Bethe lattice, there are three edges pointing outwards from $0$, but for all vertices labeled $2$, there is one edge pointing in and two edges pointing out. 
For a general degree $d\geq 3$, apart from the origin, every other vertex has one edge pointing in and $d-1$ edges pointing outwards, while the origin has $d$ edges pointing out. 
Therefore, we do not have translation invariance and we cannot shift the starting point to any other vertex and simultaneously preserve the generating functions.
Nevertheless, as we did on the line, we can show that the transition weight between two arbitrary lattice sites $z$ and $w$, is related to the transition weight from $0$ to the site $|g(z,w)|$.

\vskip 12pt
\noindent
\newpage
{\bf {{\noindent From vertex to vertex.}}}
The same discussion that we had on the line allow us to 
conclude that, for two vertices distant by an even number of steps
\be
\hat{p}_{2\t}(2x,2y) = \hat{p}_{2\t}(0, |g(2x,2y)|)\, .
\ee
We should note that, in the case of the $d \ge 3$ Bethe lattice, a vertex is connected to $d$ edges, but an edge is connected two vertices. Therefore, unlike the case of the line, we cannot relate the transition weights between two edges separated by a distance $|g(2x,2y)|$ to $\hat{p}_{2\t}(0,|g(2x,2y)|)$.
\vskip 12pt
\noindent

\begin{figure}[H]
    \centering
    \begin{minipage}[t]{.75\textwidth}
    \centering
    \includegraphics[width=7.7cm]{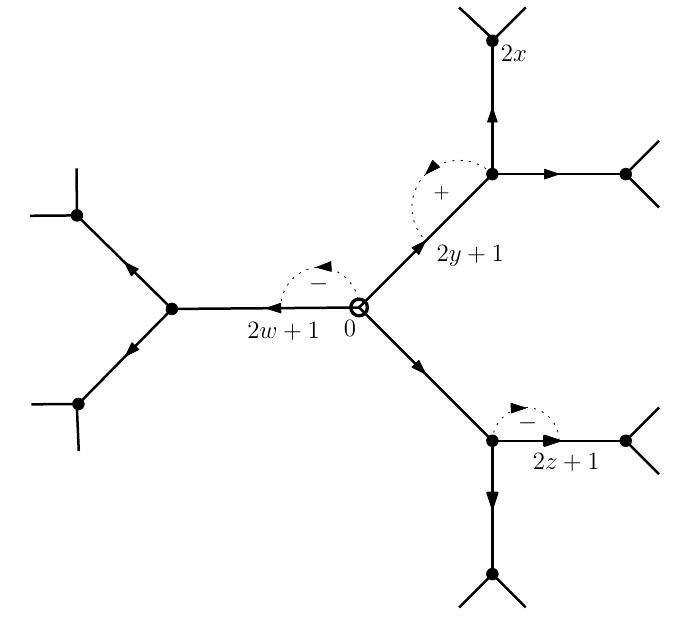}
    \caption{
    A directed Bethe lattice with an arbitrary starting vertex $2x$ for the Dirac walker. If the ending point is $2y+1 \in g(0,2x)$,  the  last step for the geodesic walk to the ending point $2y+1$ is towards the origin, hence accompanied by a $+$ sign. On the other hand, if the ending point does not lie on the geodesic $g(0, 2x)$, e.g., $2w+1$ or $2z+1$, the last step in the geodesic walk to the ending edge takes a $-$ sign.   
    }
    \label{fig:directedBetheVtoE}
    \end{minipage}
\end{figure}

{\bf {{\noindent From vertex to edge.}}}
We now consider Dirac geodesic walks from a vertex to an edge.
If we start at the origin, the last step on the geodesic is always away from the origin and thus contributes to a sign of $-1$. 
For a walk from a vertex to an edge say $2x$ to $2y+1$, the sign would be determined by the distance between the underlying vertex to vertex walk from $2x$ to the nearest vertex of the terminal edge $2y+1$ on $g(2x,2y+1)$ and the sign of the last step on the geodesic. 
In this case, the sign of the last step depends on relative positions of $2x$ and $2y+1$. The last step is towards the origin if $2y+1 \in g(0,2x)$ and otherwise it is away from the origin as shown in Figure~\ref{fig:directedBetheVtoE}. 
Therefore,
\be
\sgn(\omega_{2\t+1,2x\to 2 y+1}) =
\begin{dcases}
    -\sgn(\omega_{2\t+1,0\to |g(2x,2y+1)|}) &\text{if } 2y+1 \in g(0,2x) \,,\\
    \sgn(\omega_{2\t+1,0\to |g(2x,2y+1)|}) &\text{otherwise,}
\end{dcases}
\ee 
and for the transition weights
\be
\hat{f}_{2\t+1}(2x,2y+1) =
\begin{dcases}
    -\hat{f}_{2\t+1}(0,|g(2x,2y+1)|) &\text{if } 2y+1 \in g(0,2x) \,,\\
   \hat{f}_{2\t+1}(0,|g(2x,2y+1)|) &\text{otherwise. }
\end{dcases}
\ee 

\begin{figure}[H]
    \centering
    \begin{minipage}[t]{.75\textwidth}
    \centering
    \includegraphics[width=7.7cm]{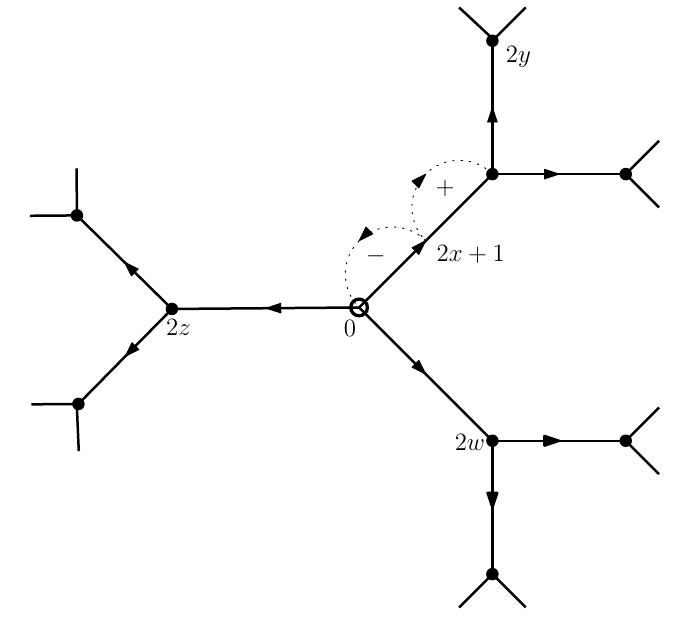}
    \caption{A directed Bethe lattice with an arbitrary starting edge $2x+1$ for the Dirac walker. If the ending point $2y$ is placed such that $2x+1 \in g(0,2y)$, the first step for the geodesic walk is away from the origin, hence has a $+$ sign. Otherwise, if the starting point does not lie on the geodesic e.g., $g(0,2z)$ or $g(0,2w)$ for the ending points $2z$ or $2w$ respectively, the first step is towards the origin, and therefore has a $-$ sign.
    }
    \label{fig:directedBetheEtoV}
    \end{minipage}
\end{figure}

\vskip 12pt
\noindent{\bf{{From edge to vertex.}}}
\noindent
We will look at walks starting from an edge $2x+1$ and ending on a vertex $2y$. 
We introduce  $\omega^{(E)}_{\t,z\to w}$, where the superscript `$(E)$' indicates that the walks are labeled in the coordinate system in which one of the edges attached to the root is considered as $0$. 
We choose to compare $\hat{p}_{2\t+1}(2x+1,2y)$ with $\hat{f}^{(E)}_{2\t+1}(0,|g(2x+1,2y)|)$. 
The lattice site $|g(2x+1,2y)|$ lies on the side of the Bethe lattice where the first step is along the orientation of the edge on the geodesic $g(0,|g(2x+1,2y)|)$ 
(e.g., on the right side in Figure~\ref{fig:edgestart})
and thus has a sign $+1$. 
The overall sign of the walk depends on the product of the signs of the first step on the geodesic and of the remaining vertex to vertex walk on the geodesic. 
If $2x+1 \in g(0,2y)$, the first step is away from the origin and has a sign $+1$, otherwise the first step is towards the origin and contributes to a sign $-1$ (cf. Figure~\ref{fig:directedBetheEtoV}).
Consequently,
\be
\label{eq:edgetoegesigngen}
\sgn(\omega_{2\t+1,2x+1\to 2 y}) =
\begin{dcases}
    \sgn(\omega^{(E)}_{2\t+1,0\to |g(2x+1,2y)|})  &\text{if } 2x+1 \in g(0,2y) \,,\\
    -\sgn(\omega^{(E)}_{2\t+1,0\to |g(2x+1,2y)|})  &\text{otherwise.}
\end{dcases}
\ee 
Therefore, the transition weights read 
\be
\hat{p}_{2\t+1}(2x+1,2y)  =
\begin{dcases}
    \hat{f}^{(E)}_{2\t+1}(0,|g(2x+1,2y)|)   &\text{if } 2x+1 \in g(0,2y) \,,\\
    -\hat{f}^{(E)}_{2\t+1}(0,|g(2x+1,2y)|)  &\text{otherwise.}
\end{dcases}
\ee 

\begin{figure}[H]
    \centering
    \begin{minipage}[t]{.75\textwidth}
    \centering
    \includegraphics[width=7.7cm]{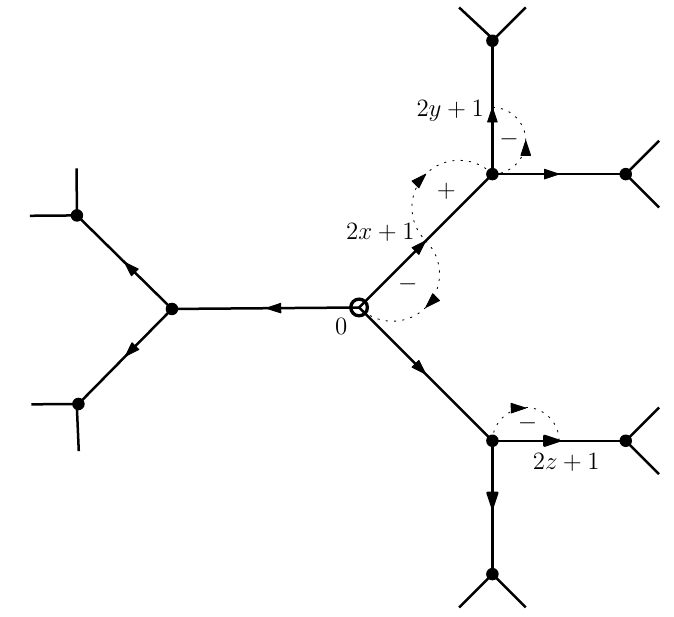}
    \caption{A directed Bethe lattice with arbitrary starting edge $2x+1$. If the ending edge is $2y+1$ such that $2x +1 \in g(0,2y+1)$, both the first and the last steps are away from the origin and they have opposite signs. Hence, their product contributes to a $-$ sign. If the ending edge is $2z+1$, such that $2x+1$ does not lie in the geodesic $g(0,2z+1)$, and $2z+1$  does not lie in $g(0,2x+1)$; for the geodesic walk, the first step from the starting edge is towards the origin and the last step to the ending edge is away from the origin on the geodesic $g(2x+1,2z+1)$. Consequently, both steps have the same signs and their product contributes to a $+$ sign.
    }
    \label{fig:directedBetheEtoE}
    \end{minipage}
\end{figure}

\vskip 12pt
\noindent
\newpage
{\bf {{\noindent From edge to edge.}}}
For an edge to edge walk, we will relate $\hat{f}_{2\t}(2x+1,2y+1)$ to $\hat{p}^{(E)}_{2\t}(0,|g(2x+1,2y+1)|)$, 
where the edge labeled $|g(2x+1,2y+1)|$ lies on the side of the Bethe lattice where the first step on the geodesics is along the orientation of the edge.
The overall sign of the walk depends on the first and the last steps and the distance between 
the first and the last vertices of the geodesic.
If $2x +1 \in g(0,2y+1)$ or $2y+1 \in g(0,2x+1)$, the first and the last steps are away from or towards the origin respectively; 
in such cases, the first and last steps contribute an overall sign of $-1$ to the walk. 
If one of the last or the first step is towards the origin and the other is away from the origin, then the steps contribute to an overall $+1$ sign to the walk.
 For any walk $\omega^{(E)}_{2\t,0\to|g(2x+1,2y+1)|}$, both the first and the last steps are away from the origin which leads to a contribution of $-1$ sign from the last and the first steps, as shown in Figure~\ref{fig:directedBetheEtoE}. Subsequently,
 \be
\sgn(\omega_{2\t,2x+1 \to 2y+1}) =
\begin{dcases}
    \sgn(\omega^{(E)}_{2\t,0\to|g(2x+1,2y+1)|})  &\text{if } 2x+1 \in g(0,2y+1) \text{ or } 2y+1 \in g(0,2x+1) \,,\\
    - \sgn(\omega^{(E)}_{2\t,0\to|g(2x+1,2y+1)|})  &\text{otherwise.}
\end{dcases}
\ee
 Therefore,
  \be
\hat{f}_{2\t}(2x+1,2y+1)=
\begin{dcases}
    \hat{p}^{(E)}_{2\t}(0,|g(2x+1,2y+1)|)  &\text{if } 2x+1 \in g(0,2y+1) \text{ or } 2y+1 \in g(0,2x+1) \,,\\
    -\hat{p}^{(E)}_{2\t}(0,|g(2x+1,2y+1)|)  &\text{otherwise.}
\end{dcases}
\ee

\vskip 10pt
\noindent
 {\bf {Summary.}}
\newline
As the arguments hold for general $\t$, we conclude that the generating functions $\hat{H}$, $\hat{P}$, $\hat{F}$, $\hat{P}^{(E)}$, and $\hat{F}^{(E)}$ obey the relations
\begin{align}
\label{eq:RWgeneral}
\hat{H}_{x,y}(\l) &= \hat{H}_{0,\abs{g(x,y)}}(\l)
\,,
\\
\label{eq:VtoVDWgeneral}
\hat{P}_{2x,2y}(\l) &= \hat{P}_{0, |g(2x,2y)|}(\l)\, ,
\\
\label{eq:VtoEDWgeneral}
\hat{F}_{2x,2y+1}(\l) &=
\begin{dcases}
    -\hat{F}_{0,|g(2x,2y+1)|}(\l) &\text{if } 2y+1 \in g(0,2x) \,,\\
   \hat{F}_{0,|g(2x,2y+1)|}(\l) &\text{otherwise.}
\end{dcases}
\\
\label{eq:EtoVDWgeneral}
\hat{P}_{2x+1,2y}(\l)  &=
\begin{dcases}
    \hat{F}^{(E)}_{0,|g(2x+1,2y)|}(\l)   &\text{if } 2x+1 \in g(0,2y) \,,\\
    -\hat{F}^{(E)}_{0,|g(2x+1,2y)|}(\l)  &\text{otherwise.}
\end{dcases}
\\ 
\label{eq:EtoEDWgeneral}
\hat{F}_{2x+1,2y+1}(\l) &=
\begin{dcases}
    \hat{P}^{(E)}_{0,|g(2x+1,2y+1)|}(\l)  &\text{if } 2x+1 \in g(0,2y+1) \text{ or } 2y+1 \in g(0,2x+1) \,,\\
    -\hat{P}^{(E)}_{0,|g(2x+1,2y+1)|}(\l)  &\text{otherwise.}
\end{dcases}
\end{align}

\subsection{Asymptotic expansion}
\label{sec:asymptote}

The spectral dimension $d_s$ of a geometry, with respect to simple random walks is determined by the power law scaling (assuming that a power law behavior exists)  of the self return probability in $\t$ steps $h_{\t}(z,z)$ in the limit $\t \to \infty$ \cite{alexander1982density,Correia_1998,Ambj_rn_2005}. If the simple random walk starts at some vertex $z$,
\be
h_{\t}(z,z) \sim \frac{C^{-\t}}{\t^{\frac{d_s}{2}}}
\,,
\ee
where $C$ is some constant such that $|C|\ge 1$ (which may have different numerical value in different equations) and the symbol $\sim$ represents asymptotic scaling at large $\tau$.
\footnote{By asymptotic scaling, we mean that the quantities on the left and right of the symbol differ only by a multiplicative constant in the appropriate limit, and the constant is independent of the limiting variable.}
Analogously, we wish to generalize the definition to self return transition weights for the Dirac walk, and define the spectral dimension of a geometry with respect to the Dirac walk starting at some vertex $2z$ as
\be
\label{eq:DiracSpectral}
p_{\t}(2z,2z) \sim \frac{C^{-\t}}{\t^{\frac{d_s}{2}}}
\,,
\ee
and similarly for edges, using the transition weight $f_\tau$.
We will derive the power law scaling of self return transition weights for the Dirac walks and probabilities for simple random walks using asymptotic analysis of the generating functions.
In the context of asymptotic analysis, it is useful to view our generating functions as complex-valued functions in $\l$. Our strategy would be to use singularity analysis described in \cite{10.5555/1506267}. It was shown that for a function of type $(b-a \l)^{-\a}$~, 
where $\a$ is a real number barring negative integers ($\a \notin \mathbb{Z}_{-}$), 
having a dominant singularity at $\frac{b}{a}$,
\beq
\label{eq:standardfunction}
\left[\l^{\t}\right](b-a\l)^{-\alpha} \sim \left(\frac{a}{b}\right)^{\t}\frac{\t^{\alpha-1}}{\Gamma(\alpha)}\left(1+\mathcal{O}\left(\frac{1}{\t}\right)\right)\, .
\eeq
The symbol $\mathcal{O}(\vert f(x)\vert)$ represents terms that are bounded by  
 $c |f(x)|$
for some positive constant $c$, and the singularity closest to $0$ is called the dominant singularity. 
$\left[\l\right]^{\t}$ represents the $\t^{\text{th}}$ coefficient in series expansion of $(b-a \l)^{-\a}$.

\subsubsection{Asymptotics of walks on the line}

On the line (or the $d=2$ Bethe lattice), for both the Dirac and the simple random walk, $H_{0,0}(\l)= P_{0,0}(\l)$. From Equations \eqref{eq:GFforlineSRW0} and \eqref{eq:PxVtoVline}, we recall that the generating functions for self return probability (weights) are
\beq
P_{0,0}(\l)=H_{0,0}(\l) =  \frac{1}{\sqrt{1-\lambda ^2}}\, .
\eeq
We are interested in coefficients of $\l^{2\t}$ as odd powers of $\l$ vanish,
since it takes an even number of steps to return back to itself.
By definition of generating functions and a direct application of Equation \eqref{eq:standardfunction}, with $a=1$, $b=1$ and $\a = \frac{1}{2}$, we get
\beq
h_{2\t}(0,0) = [\l^{2\t}]H_{0,0}(\l) \sim \frac{1}{(\pi \t)^{\frac{1}{2}}}\, .
\eeq
Therefore, we conclude that on the line, both the unbiased 
Dirac walk $(\alpha = \beta = 1/2)$ and the simple random walk $(\alpha = 1/2)$ have a spectral dimension $d_s = 1$.

\subsubsection{Asymptotics of walks on Bethe lattice}

{\bf {Simple random walk.}}

We recall that for $d \geq 3$ Bethe lattice,  as given in \eqref{eq:PxVtoVline},
To perform the singularity analysis of \eqref{eq:standardfunction}, we need a series expansion around the dominant singularity using the methods described in \cite{10.5555/1506267}. 
The dominant singularity for $\hat{H}_{0,0}(\l)$ lies at 
$\l^2 = \frac{d^2}{4(d-1)} $ and the series expansion around it reads 
\be
\hat{H}_{0,0}(\l) \simeq
K(d)
-
G(d)
\sqrt{\frac{d^2}{4 (d-1)}-\lambda ^2}
+\mathcal{O}\left(\left(\frac{d^2}{4 (d-1)}-\lambda ^2\right)^{3/2}\right)
\,,
\ee
where $K(d)$ and $G(d)$ are functions of $d$, independent of $\l$. 
Applying \eqref{eq:standardfunction} to the leading order term which depends on $\l^2$ with $a=1$, $b =  \frac{d^2}{4(d-1)}$ and $\a = \frac{-1}{2}$ obtain
\beq
\hat{h}_{2\t}(0,0) = [\l^{2\t}] \hat{H}_{0,0}(\l) \sim \frac{\left(\frac{d^2}{4(d-1)}\right)^{-\t}}{\sqrt{\pi}(\t)^{\frac{3}{2}}}
\,.
\eeq
We conclude that $d_s = 3$ for a simple random walk, for any $d$, agreeing with the result of \cite{monthus1996random}. 

\vskip 5pt
\noindent
{\bf{Dirac walk.}}
We are interested in coefficients of $\l^{2\t}$ as odd powers of $\l$ are identically $0$. One can see that in \eqref{eq:VertexStartingDiracBethe}, the dominant singularity is at  $\lambda ^2=\frac{2 d}{d+2 \sqrt{d-1}} $. 
A bit of algebra reveals that
\beq
\hat{P}_{0,0}(\l) \simeq A(d)+ C(d)\sqrt{ \frac{2 d}{d+2 \sqrt{d-1}}-\l^2}+\mathcal{O}\left(\frac{2 d}{d+2 \sqrt{d-1}}-\lambda ^2\right) \, ,
\eeq
where $A(d)$ and $C(d)$ are functions of $d$. The explicit form is not written to reduce clutter in the expressions.
Therefore one obtains by \eqref{eq:standardfunction} with $a=1$, $b = \frac{2 d}{d+2 \sqrt{d-1}}$ and $\a = \frac{-1}{2}$
\beq
 \hat{p}_{2\t}(0,0) = [\l^{2\t}]\hat{P}_{0,0}(\l) \sim \frac{\left(\frac{2 d}{d+2 \sqrt{d-1}}\right)^{-\t}}{\sqrt{\pi}\t^{\frac{3}{2}}}
 \,,
\eeq
and like the case of the line, the Dirac walk on the Bethe lattice sees $d_s = 3$ on the vertex, as does the simple random walk.

Similarly from \eqref{eq:EdgetoEdgehat},
the dominant singularity lies again at $\lambda ^2=\frac{2 d}{d+2 \sqrt{d-1}}$ and 
\beq
\hat{P}^{(E)}_{0,0}(\l) \simeq A^{(E)}(d)+ C^{(E)}(d)\sqrt{ \frac{2 d}{d+2 \sqrt{d-1}}-\l^2}+\mathcal{O}\left(\frac{2 d}{d+2 \sqrt{d-1}}-\lambda ^2\right) \, ,
\eeq
where $A^{(E)}(d)$ and $C^{(E)}(d)$ are functions of $d$. 
Therefore one obtains by \eqref{eq:standardfunction}  with $a=1$, $b = \frac{2 d}{d+2 \sqrt{d-1}}$ and $\a = \frac{-1}{2}$,
\beq
 \hat{p}^{(E)}_{2\t}(0,0) = [\l^{2\t}]\hat{P}^{(E)}_{0,0}(\l) \sim \frac{\left(\frac{2 d}{d+2 \sqrt{d-1}}\right)^{-\t}}{\sqrt{\pi}(\t)^{\frac{3}{2}}}\, ,
\eeq
which implies that the Dirac walk on the Bethe lattice also sees a spectral dimension $d_s = 3$ on edges.

\section{QFT on the Bethe lattice}
\label{ch-5}

In this section, we combine the results obtained in Section \ref{ch-1} to discuss the two-point function $C(x,y)$ obtained for free scalars in \eqref{eq:resolventgeneralscalar} and free Fermions in \eqref{eq:ResolventgeneralDirac} on the Bethe lattice.
The analytic properties of the Green function of an operator contain information about its spectrum. For details, one can see \cite{economou2006green}, and some relevant results have been reviewed in Section \ref{ssec:GreensFunction}. 
After recalling the Green function of free scalars, we will present our results for free Fermions on the Bethe lattice.

\subsection{Brief overview of Green function}{\label{ssec:GreensFunction}}
The Green function $G(m)$ of a linear Hermitian differential operator $L$ is defined as the resolvent with respect to some complex-valued parameter $m$. In operator form, it can be written as
\begin{equation}
\label{eq:DefG}
G(m) = \frac{1}{m - L}
\end{equation}
and in terms of matrix elements,
\begin{equation}
[m-L(x)] G(x, y; m) = \delta(x - y). \nonumber
\end{equation}
Simple real poles correspond to discrete eigenvalues of $L$, while branch cuts on the real $m$-axis correspond to continuous eigenvalues. It is worth noting that there could be other types of singularities in $G(m)$ \cite{economou2006green}, but that discussion is beyond our scope. If $m = E$, where $E$ belongs to the continuous spectrum, it is useful to define the following Green functions, analytic in the upper and lower half-plane respectively
\begin{align}
G^{+}(x, y; E) &:= \lim_{s \rightarrow 0^{+}} G(x, y; E + \mathrm{i} s)\,,  \\
G^{-}(x, y; E) &:= \lim_{s \rightarrow 0^{+}} G(x, y; E - \mathrm{i} s)\,.  
\end{align}
The density of states per unit volume is given by the discontinuity across the cut
\begin{equation}
\varrho(x; E) = \mp \frac{1}{\pi} \mathrm{Im}{G^{\pm}(x, x; E)}
=
\abs{\frac{1}{\pi} \mathrm{Im}{G^{\pm}(x, x; E)}}.\label{eq:DOS}
\end{equation}
Detailed derivations can be found in the book \cite{economou2006green}.

\subsection{Scalar QFT on the Bethe lattice}
\label{sec:SFT}
We will review here the Green functions and spectra of scalar fields on the Bethe lattice, first $d=2$, corresponding to the line, then $d\geq 3$, retrieving results from \cite{economou2006green}.

We recall that for the scalar field theory on a graph, by setting $L = - \Delta_\Gamma$ which is the negative graph Laplacian in \eqref{eq:DefG}, the Green function takes the form
\be
\label{eq:resolventgeneralscalar2}
G(m) = \frac{1}{\DG+m\mathbf{1}} = \frac{1}{D-A+m\mathbf{1}} = \frac{1}{D+m\mathbf{1}}\sum^{\infty}_{\t=0}\left(\frac{1}{D+m\mathbf{1}}A\right)^{\t} \,.
\ee
As the Bethe lattice is $d$-regular, the degree matrix takes a rather simple form $D = d \mathbf{1}$. Therefore, we can write 
\be
\label{eq:resolventgeneralscalar2b}
G(m) = \frac{1}{d+m}\sum^{\infty}_{\t=0}\left(\frac{A}{d+m}\right)^{\t} \,.
\ee
Reparametrizing $m = \frac{d(1-\l)}{\l}$, we get in the sum of \eqref{eq:resolventgeneralscalar2} for each matrix element
\be
\sum^{\infty}_{\t=0}
\left(
\frac{A}{d+m}
\right)^{\t}(x,y)
= 
\sum^{\infty}_{\t=0}
\l^{\t}\left(
\frac{A}{d}\right
)^{\t}(x,y) 
 \,.
\ee
The right-hand side is the generating function $H_{x,y}(\l)$ for the line and $\hat{H}_{x,y}(\l)$ for $d \ge 3$ Bethe lattice respectively, as discussed in Section~\ref{sec:SFTonGraphs}.
 
\subsubsection{Free scalar on the line}

\begin{figure}[H]
    \centering
    \begin{minipage}[t]{.8\textwidth}
    \centering
    \includegraphics[width = 13cm]{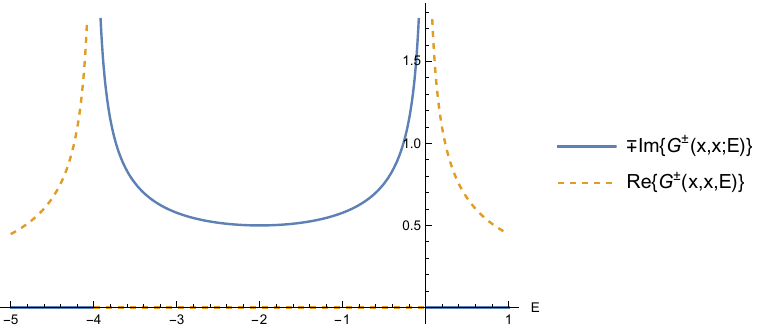}
    \caption{Plot of the autocorrelation $G(x,x;E)$. The blue line is $\mp {\rm Im} \{{G^{\pm}(x,x;E)}\} $ and the orange dashed line is ${\rm Re}\big\{{G^\pm(x,x;E)}\big\}$.}
    \label{fig:DOS for 1D free scalar}
    \end{minipage}
\end{figure}

For a line, one can compute the Green function by setting 
$\l = \f{d}{d+m}= \frac{2}{2+m}$  and using \eqref{eq:GFforlineSRW0},
\beq
\label{eq:Greens1D}
 G(x,y;m)
 = 
 \frac{
 H_{x,y}(2/(2 + m))}{2 + m}
 =\frac{
 \left(
 \frac{1}
 {2}
 \left(
 m+2
 -
 \sqrt{
 m (m+4)
 }\right)
\right)^{|g(x,y)|}}
{
\sqrt
{m (m+4)
}
} \,.
\eeq
We look at the auto-correlation function
\be
G(x,x;m) =
\frac{
H_{x,x}(2/(2 + m))
}
{
2 + m
} 
=
\frac{
1
}
{
\sqrt{m(m+4)
}
} \,.
\ee
The function $G(x,x;m)$ has a branch cut between $-4<m<0$. If $m = E \pm i s$, where $E$ is an eigenvalue of our (negative) Laplacian, we compute the density of states per site $\varrho(x;E)$ as
\beq 
\label{eq:1DDOS}
\varrho(x;E) 
= 
\mp \frac{1}{\pi} {\rm Im} \{ G^{\pm}(x,x;E)\}
= 
\frac{1}{\pi\sqrt{-E(E+4)}}
\,,
\eeq
where $E \in (-4,0)$ as explained in \eqref{eq:DOS}.
The branch points, denoted by  $E_g$, are called the band edges of the Green function. 
The behavior of $G^{\pm}(x,x;E)$ around the band edge characterizes the singularities of the Green function. For a free scalar field on a line, $E_g = 0$ or $E_g = -4$, and ${\rm Im}\{{\varrho(x;E)}\}$ diverges as $|E-E_g|^{-\f{1}{2}}$ approaching from inside the spectrum. This divergence is characteristic of one-dimensional systems \cite{economou2006green}. We observe discontinuities for both the real and imaginary part of $G(x,x;m)$ at the band edges.
The real and imaginary parts of $G^{\pm}(x,x;m)$ are plotted in Figure~\ref{fig:DOS for 1D free scalar}. 

\subsubsection{Free scalar on the $d \ge 3$ Bethe lattice}
{\label{ssec:scalarQFTonBethe}}

\begin{figure}[H]
\centering
    \begin{minipage}[t]{0.9\textwidth}
    \centering
    \begin{subfigure}[c]{0.52\linewidth}
   $\vcenter{\hbox{\includegraphics[height =4.8cm]{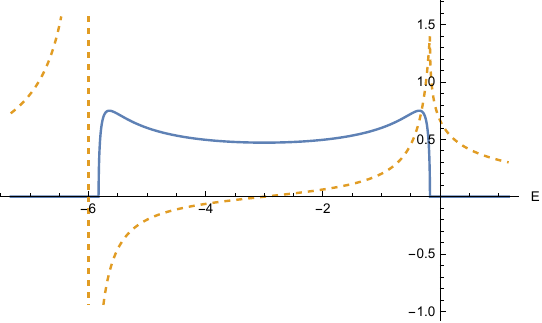}}}$
    \end{subfigure}
    \quad
    \begin{subfigure}[c]{0.42\linewidth}
$\vcenter{\hbox{\includegraphics[trim= 1.1cm 0.0cm 3.47cm 0.0 cm,clip,
    height =4.cm]{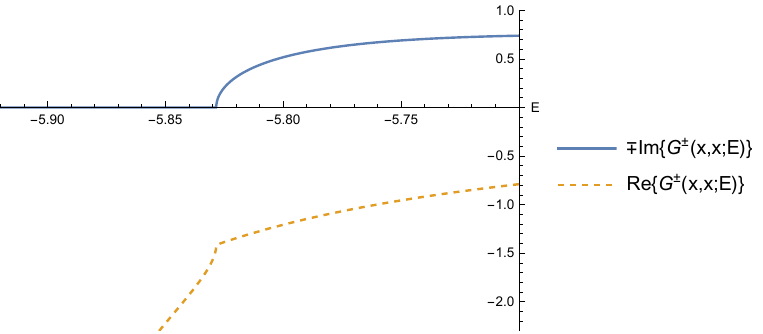}}}$
    \end{subfigure}
    \caption{
    Plot of the autocorrelation $G(x,x;m)$. The blue line is $\mp{\rm Im}\{G^{\pm}(x,x;E)\}$ and the orange dashed line is  ${\rm Re}\{G^\pm(x,x;m)\}$ for the $d=3$ Bethe lattice. ${\rm Re}\{G^\pm(x,x;m)\}$ has Van Hove singularities on the band edges $\pm2\sqrt{d-1}-d$. A zoomed in picture for the Van Hove singularity at $-2\sqrt{d-1}-d$ is shown on the right.}
    \label{fig:DOSforbethefreescalar}
    \end{minipage}
\end{figure}

Similarly to the case of line \eqref{eq:Greens1D}, we can compute the generating function by setting $\lambda = \frac{d}{d+m}$ 
and 
using \eqref{eq:RWgeneral} and \eqref{eq:singlesiteBetheRW},
\be 
 G(x,y;m)
 = 
 \frac{\hat{H}_{x,y}(d/(d + m))}{d + m}
 =
 \frac{
 2 (d-1) 
 \left(
 \frac
 {
 \left((d+m)-\sqrt{(d+m)^2-4 (d-1)}
 \right)
 }
 {
 2 (d-1)
 }
 \right)^{|g(x,y)|}}
 {
 \left(d \sqrt{(d+m)^2-4 (d-1)}+(d-2)(d+m)
 \right)
 }\,,
\ee
and
\beq
G(x,x;m) =
\frac{\hat{H}_{x,x}(d/(d + m))}{d + m}=\frac{2 (d-1)}{\left(d \sqrt{(d+m)^2-4 (d-1)}+(d-2)(d+m)\right)}\,.
\eeq
Clearly $G(x,x;m)$ has a branch cut for $m \in (-2 \sqrt{d-1}-d,2 \sqrt{d-1}-d)$ and a simple pole at $m = -2d$. This gives us the range of eigenvalues of the graph Laplacian of the Bethe lattice. We can then deduce the density of states per unit volume 
\beq
\varrho(x;E)= \mp\frac{1}{\pi} {\rm Im}\{G^{\pm}(x,x;E)\}=\frac{d\sqrt{-(d+E)^2+4 (d-1)}}{\pi \left(4 d E+2 E^2\right)},
\eeq
for $E \in (-2 \sqrt{d-1}-d,2 \sqrt{d-1}-d)$. 
Close to the band edges $E_g=\pm2 \sqrt{d-1}-d$, the density $\varrho(x;E)$ approaches zero as $ |E-E_g|^{\f{1}{2}}$ when $E \to E_g$ from inside the spectrum for both band edges. 
Except for the pole at $E=-2d$, the real and imaginary parts of $G^{\pm}(x,x;E)$ are continuous. 
There are Van Hove singularities at the edges of the spectrum.
The real and imaginary parts of $G^{\pm}(x,x;E)$ are plotted in Figure \ref{fig:DOSforbethefreescalar}.

\subsection{Fermionic QFT on the Bethe lattice}

We recall from \eqref{eq:ResolventgeneralDirac} that 
\beq
G_{\sD}(z,w;m)=\left(\frac{1}{-\slashed{D}_{I}+m \mathbf{1}}\right)(z,w)=\frac{1}{m} \sum_{\t=0}^{\infty}\left(\frac{\slashed{D}}{m}\right)^{\t}(z,w) \,.\label{eq:ResolventgeneralDirac2}
\eeq
In this section we will relate the powers of Dirac operator $\sD$ to the transition weights computed in Section \ref{ch-2} and combinatorially compute the Green function for the free Fermion on graph. Analogous to Section
\ref{sec:SFT}, we will deduce the spectrum of the Dirac operator $\sD$ on the line ($d=2$ Bethe lattice) and the  $d \ge 3$ Bethe lattice.

\subsubsection{Fermionic QFT on the line}

\begin{figure}[H]
    \centering
    \begin{minipage}[t]{.8\textwidth}
    \centering
    \includegraphics[width = 13cm]{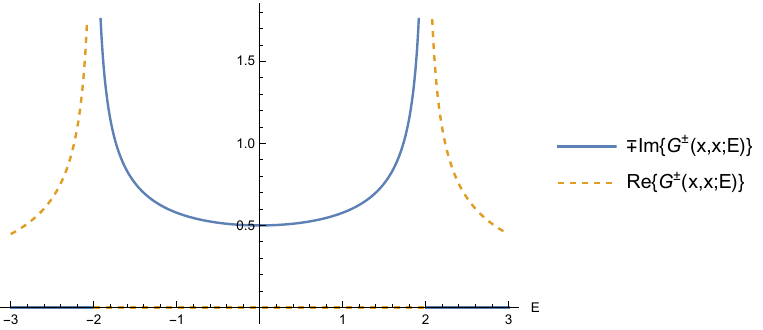}
    \caption{
    Plot of $G_{\sD}(x,x;E)$. The blue line is 
    $\mp {\rm Im}\{{G_{\sD}^{\pm}(x,x;E)}\}$ and the orange dashed line is 
    ${\rm Re}\{{G_{\sD}^{\pm}(x,x;E)}\}$. 
    }
    \label{fig:DOS for 1D free fermion}
    \end{minipage}
\end{figure}

\noindent
{\bf {Vertex to vertex or edge to edge.}}
We choose two vertices $2x$ and $2y$ and two edges $2z+1$ and $2w+1$ such that they are separated by the same distance $|g(2x,2y)| = |g(2z+1,2w+1)|=2\gamma$.
Recall also that by definition, the even powers of the Dirac operator are given by $\sD^{2\t}(2x,2y) = 2^{2\t}p_{2\t}(2x,2y)$ between vertices and similarly $\sD^{2\t}(2z+1,2w+1) = 2^{2\t}f_{2\t}(2z+1,2w+1)$ between edges.
Therefore,
from \eqref{eq:VtoVEtoEDWgeneral}
\begin{gather}
    G_{\sD}(2x,2y;m)=\f{1}{m}
P_{2x,2y}\left(\f{2}{m}\right)=P_{0,2\g}(\l)\,,\\
    G_{\sD}(2z+1,2w+1;m)=\frac{1}{m}
    \,,
F_{2z+1,2w+1}\left(\f{2}{m}\right)=P_{0,2\g}(\l)
\,.
\end{gather}
From \eqref{eq:VtoVEtoEDWgeneral} and \eqref{eq:PxVtoVline} we obtain explicitly 
\be
G_{\sD}(2x,2y;m)=G_{\sD}(2z+1,2w+1;m)
= 
\frac{\left(\frac{1}{2} m \left(\sqrt{m^2-4}-m\right)+1\right)^{\g }}{\sqrt{m^2-4}}
\,.
\ee
For any lattice site $x$, we find the autocorrelation 
\be
G_{\sD}(x,x;m)= \frac{1}{\sqrt{m^2-4}}
\, ,
\ee
that has a branch cut for $-2<m<2$. This gives us the range of eigenvalues of $\sD$ on the line.
From $G_{\slashed{D}}(x,x;E)$, where $E \in (-2,2)$, we compute the density of states per site
$\varrho_{\sD}(x;E)$
as
\beq \label{eq:1DfreeFermionDOS}
\varrho_{\slashed{D}}(x;E) 
= 
\mp \frac{1}{\pi} {\rm Im} 
\big\{{G_{\slashed{D}}^{\pm}(x,x;E)}\big\}
= 
\frac{1}{\pi \sqrt{(4-E^2)}} \,.
\eeq
One sees that eigenvalues of $\sD$ are the square root of the eigenvalues of the Laplacian. Exactly like the case of free scalar in \eqref{eq:1DDOS}, the density of states for free Fermions in \eqref{eq:1DfreeFermionDOS} diverges as $|E-E_g|^{-\f{1}{2}}$ as $E \to E_g$ on the branch cut. Like the case of the scalar QFT, we observe discontinuities for the real and imaginary parts of auto-correlation function at the band edges. The real and imaginary part of $G_{\sD}^{\pm}(x,x;E)$ is plotted in Figure~\ref{fig:DOS for 1D free fermion}.

\noindent
{\bf {Vertex to edge.}}
By definition, the odd powers of the Dirac operator rewrite as
$\sD^{2\t+1}(2x,2y+1) = 2^{2\t+1}f_{2\t+1}(2x,2y+1)$.
Therefore, the Green function is
\be
G_{\sD}(2x,2y+1;m)= \frac{1}{m} F_{2x,2y+1}\left(\frac{2}{m}\right)
\, .
\ee
\\
If the sites are at distance $\abs{g(2x,2y+1)}=2\d+1$, we find explicitly from \eqref{eq:VtoElinegeneral} and \eqref{eq:FxVtoEline}
\begin{gather}
\label{eq:VtoEDiracGreenBethe}
\mathcal{G}(\d,m)=\frac{\left(\sqrt{m^2-4}-m\right) \left(\frac{1}{2} \left(m \left(\sqrt{m^2-4}-m\right)+2\right)\right)^{\d}}{2 \sqrt{m^2-4}}
\,,
\\
G_{\sD}(2x,2y+1;m) =
\begin{dcases}
   \mathcal{G}(\d,m)
   &\text{if } 2x<2y+1
   \, ,
   \\
  -\mathcal{G}(\d,m) &\text{otherwise.}
\end{dcases}
\end{gather}
\noindent
{\bf {Edge to vertex.}}
As above, the odd powers of the Dirac operator relate to weighted sum of walks as 
$\sD^{2\t+1}(2y+1,2x) = 2^{2\t+1}f_{2\t+1}(2y+1,2x)$.
Therefore, the Green function is given by
\be
G_{\sD}(2y+1,2x;m)= \frac{1}{m} P_{2y+1,2x}\left(\frac{2}{m}\right)
\,
.
\ee
\\
If the sites are at distance $|g(2y+1,2x)|= 2\d+1$, we obtain from \eqref{eq:EtoVGeneral} and \eqref{eq:FxVtoEline} and 
\be
\label{eq:EtoVDiracGreenBethe}
G_{\sD}(2y+1,2x;m) =
\begin{dcases}
   -\mathcal{G}(\d,m)
   &\text{if } 2y+1 <2x
   \, ,
   \\
  \mathcal{G}(\d,m) &\text{otherwise.}
\end{dcases}
\ee

\subsubsection{Fermionic QFT on the $d \ge 3$ Bethe lattice}

The Green functions for Fermionic QFT on Bethe lattice boil down to four distinct cases below.

\begin{figure}[H]
    \centering
    \begin{minipage}[t]{.8\textwidth}
    \centering
    \includegraphics[width = 14cm]{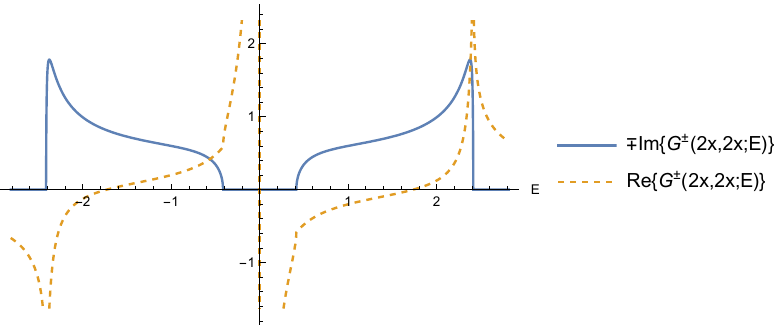}
    \caption{Plot of $G_{\sD}(2x,2x;E)$. The blue line is $\mp{\rm Im}\{G^{\pm}(2x,2x;E)\}$ and the orange dashed line is  ${\rm Re}\{G^\pm(2x,2x;E)\}$ for $d=3$ Bethe lattice. We observe a simple pole at $E=0$ and Van Hove singularities on the band edges.
    } 
    \label{fig:DOS for bether free fermion}
    \end{minipage}
\end{figure}

\vskip 5pt
\noindent
{\bf {Vertex to vertex.} }
In contrast with the line, the even powers of the Dirac operator contain a factor $d$, $\sD^{2\t}(2x,2y) = (2d)^{\t} \hat{p}_{2\t}(2x,2y)$\footnote{$\sD^{2\t+1}(2x,2y) = 0$ identically, because we cannot reach a vertex from a vertex in an odd number of steps.},
entailing
\al{
G_{\sD}(2x,2y;m) 
&=
\frac{1}{m} 
\sum^{\infty}_{\t=0} \biggl(
\frac{\sD}{m}
\biggr)^{2\t}(2x,2y)\nnn
&=\frac{1}{m} 
\sum_{\tau=0}^{\infty}\left(
\frac{\sqrt{2 d}}{m}
\right)^{2 \tau} 
\hat{p}_{2 \tau}(2 x, 2 y) 
=
\frac{1}{m}
\hat{P}_{2 x, 2 y}\left(\frac{\sqrt{2 d}}{m}\right)
\, ,}
Explicitly from \eqref{eq:VtoVDWgeneral} and \eqref{eq:VertexStartingDiracBethe}, we obtain
\be \label{eq:GreenVertexSelfBetheDirac}
G_{\sD}(2x,2y;m) =
-\frac{2 (d-1) m \left(\frac{\sqrt{-2 d m^2+(d-2)^2+m^4}+d-m^2}{2 (d-1)}\right)^{\frac{1}{2} | g(2 x,2 y)| }}{d^2-d \left(\sqrt{-2 d m^2+(d-2)^2+m^4}+m^2+2\right)+2 m^2}
\, .
\ee
To compute the density of states per vertex, we have the autocorrelation
\be
G_{\sD}(2x,2x;m) = 
-\frac{2 (d-1) m}{d^2-d \left(\sqrt{-2 d m^2+(d-2)^2+m^4}+m^2+2\right)+2 m^2}\,.
\ee
Clearly, $G_{\sD}(2x,2x;m)$ has two branch cuts for $m \in (-\sqrt{d+2 \sqrt{d-1}},-\sqrt{d-2 \sqrt{d-1}})$ and  $m\in(\sqrt{d-2 \sqrt{d-1}},\sqrt{d+2 \sqrt{d-1}})$. There is a simple pole at $m = 0$. This gives us the range of eigenvalues of the Dirac operator on the Bethe lattice. We can then deduce the density of states per site
\be
\varrho_{\sD}(2x;E) 
= \frac{d \sqrt{2 d E^2-(d-2)^2-E^4}}{2\pi E\left(E^2-2 d\right)} ,
\ee
for $E \in (-\sqrt{d+2 \sqrt{d-1}},-\sqrt{d-2 \sqrt{d-1}}) \bigcup(\sqrt{d-2 \sqrt{d-1}},\sqrt{d+2 \sqrt{d-1}})$. This time, we have two bands, separated by a gap. Exactly like for the free scalar, the density of states approaches zero as $|E-E_g|^{\f{1}{2}}$ from inside the spectrum, with Van Hove singularities at the four band edges. 
The real and imaginary parts of an example Green function are plotted in Figure~\ref{fig:DOS for bether free fermion}.

\vskip 10pt
\noindent
{\bf {Vertex to edge.}}
Odd powers of the Dirac operator, $\sD^{2\t+1}(2x,2y+1) = d (2d)^{\t} \hat{f}_{2\t+1}(2x,2y+1)$, imply the relation
\be
 G_{\sD}(2x,2y+1;m)
 =
  \frac{1}{m}
 \sqrt{\f{d}{2}}
\hat{F}_{2x,2y+1}\left(\frac{\sqrt{2d}}{m}\right)
 \,.
\ee
From \eqref{eq:VtoEDWgeneral} and \eqref{eq:VertexStartingDiracBetheF}, we read for sites are at distance $\abs{g(2x,2y+1)}=2\d+1$,
\begin{gather}
    \mathcal{G}_d(\d,m)=\frac{2 \left(\frac{\sqrt{-2 d m^2+(d-2)^2+m^4}+d-m^2}{2 (d-1)}\right)^{\d}}{\sqrt{-2 d m^2+(d-2)^2+m^4}-d+m^2-2}
    \,,
    \\
    G_{\sD}(2x,2y+1;m) =
\begin{dcases}
   -\mathcal{G}_d(\d,m)
   &\text{if } 2y+1 \in g(0,2x)
   \, ,
   \\
   \mathcal{G}_d(\d,m) &\text{otherwise.}
\end{dcases}
\end{gather}
Clearly, $2x \neq 2y+1$ for integer values of $x$ and $y$, therefore the question of density of states per site does not arise.

\vskip 10pt
\noindent
{\bf {Edge to vertex.}}
Between edges and vertices, odd powers of the Dirac operator 
$\sD^{2 \tau+1}(2 y+1,2 x)=2(2 d)^\tau \hat{p}_{2 \t+1}(2 y+1,2 x)$, bring about
\be
G_{\sD}(2y+1,2x;m)
=
\frac{1}{m} 
\sqrt{
\frac{2}{d}
}
\hat{P}_{2 y+1,2 x}\left(\tfrac{\sqrt{2d}}{m}\right)
\,.
\ee
From \eqref{eq:EtoVDWgeneral} and \eqref{eq:EdgetoVertexhat}, it follows
\be 
\label{eq:EtoVDiracGreenBethe2}
G_{\sD}(2y+1,2x;m) =
\begin{dcases}
  \mathcal{G}_d(\d,m)  &\text{if } 2y+1 \in g(0,2x)
   \, ,
   \\
 -\mathcal{G}_d(\d,m) &\text{otherwise.}
\end{dcases}
\ee
Similarly to the case of vertex to edge, we cannot comment on the density of states per site. 
From \eqref{eq:VtoEDiracGreenBethe} and \eqref{eq:EtoVDiracGreenBethe}, we see that the Green function is symmetric under exchange of edge and vertex position
\be
G_{\sD}(2x,2y+1;m) = G_{\sD}(2y+1,2x;m)
\,.
\ee
Notice that this symmetry is not related to the antisymmetry related to the anticommutation relations of our fields.

\begin{figure}[H]
    \centering
    \begin{minipage}[t]{.8\textwidth}
    \centering
    \includegraphics[width = 14cm]{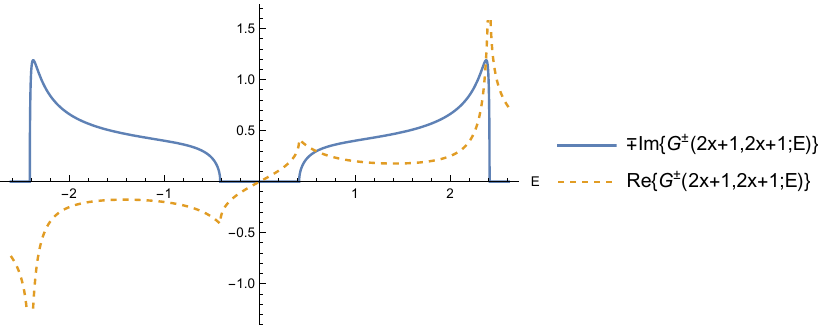}
    \caption{Plot of $G_{\sD}(2x+1,2x+1;E)$. The blue line is $\mp {\rm Im}\{G_{\sD}^{\pm}(2x+1,2x+1;E)\}$ and the orange dashed line is  ${\rm Re}\{G_{\sD}^\pm(2x+1,2x+1;E)\}$ for $d=3$ and $2x+1$ being an edge. We observe Van Hove singularities at the band edges. }
    \label{fig:DOS for bether free fermion Edge}
    \end{minipage}
\end{figure}

\vskip 10pt
\noindent
{\bf {Edge to edge.}}
Finally, even powers of the Dirac operator is $\sD^{2\t}(2x+1,2y+1) = (2d)^{\t} \hat{f}_{2\t}(2x+1,2y+1)$.
Therefore,
\be
G_{\sD}(2x+1,2y+1;m) 
=
\frac{1}{m}
\hat{F}_{2 x+1, 2 y+1}\left(\frac{\sqrt{2 d}}{m}\right)
\,.
\ee
From \eqref{eq:EtoEDWgeneral} and \eqref{eq:EdgetoEdgehat},
\begin{gather}
G_{\sD}(2x+1,2y+1;m) \hspace{10cm}
\\
\hspace{1cm}=
\begin{dcases}
\frac{m \left(\frac{\sqrt{-2 d m^2+(d-2)^2+m^4}+d-m^2}{2 (d-1)}\right)^{\frac{1}{2} | g(2 x+1,2 y+1)| }}{\sqrt{-2 d m^2+(d-2)^2+m^4}+d-2}
& \begin{array}{l} \text{if } 2x+1 \in g(0,2y+1)\text{ or }\\ 2y+1 \in g(0,2x+1)\,, \end{array}
\\
- \frac{m \left(\frac{\sqrt{-2 d m^2+(d-2)^2+m^4}+d-m^2}{2 (d-1)}\right)^{\frac{1}{2} | g(2 x+1,2 y+1)| }}{\sqrt{-2 d m^2+(d-2)^2+m^4}+d-2} 
&\text{otherwise.}
\end{dcases}
\end{gather}
The edge autocorrelation reads
\be \label{eq:Edgeself}
G_{\sD}(2x+1,2x+1;m) = \frac{m}{\sqrt{-2 d m^2+(d-2)^2+m^4}+d-2}
\,.
\ee
It has the same branch cuts as the vertex to vertex case \eqref{eq:GreenVertexSelfBetheDirac}, 
$$
m \in \left(-\sqrt{d+2 \sqrt{d-1}},-\sqrt{d-2 \sqrt{d-1}})
\bigcup
\sqrt{d-2 \sqrt{d-1}},\sqrt{d+2 \sqrt{d-1}}\right)\, ,
$$
but does not contain the pole at $ m = 0$. The edge spectral density is 
\be
\varrho_{\sD}(2x+1;E) = \mp\frac{1}{\pi} {\rm Im}\{G_{\sD}^{\pm}(2x+1,2x+1;E)\}
= 
\frac{\sqrt{-d^2+2 d \left(E^2+2\right)-E^4-4}}
{\pi E(E^2-2d)}
\,,
\ee
for $E \in (-\sqrt{d+2 \sqrt{d-1}},-\sqrt{d-2 \sqrt{d-1}}) \bigcup(\sqrt{d-2 \sqrt{d-1}},\sqrt{d+2 \sqrt{d-1}})$. Exactly like for the free scalar, the $\varrho_{\sD}(2x+1;E)$ approaches zero as $|E-E_g|^{\f{1}{2}}$ from inside the spectrum. We encounter Van Hove singularities at all four band edges. 
The real and imaginary parts of an example $G_{\sD}(2x+1,2x+1;m)$ are plotted in Figure~\ref{fig:DOS for bether free fermion Edge}. 

\section{Conclusions}
\label{sec:concl}
To summarize, our work studied various spectral properties of a random process diffusing on the Bethe lattice of degree $d\geq 2$, with the square root of the Laplacian, analogous to a Dirac operator. 
We have in particular obtained its spectral dimension as well as the Green functions between any two sites of the lattice. 
We relied on the combinatorial reformulation of the Dirac operator in terms of signed random walks suggested by \cite{casiday2022laplace} for which we computed the generating functions through recursion relations. In order to study Dirac walks starting at a vertex, we used the well-known mapping of the Bethe lattice to the half-line \cite{hughes1982random}. Here, because of oriented edges, walks starting from an edge needed a mapping to a line with a defect orientation. 
The spectral dimensions of the Dirac walk match those of the simple random walk, that is, $d_s=1$ when $d= 2$ and $d_s=3$ when $d\geq 3$. The spectra differ from those of the Laplacian by developing a gap. Although we computed the Dirac operator spectra for a fixed orientation, it is well-known that they are independent of orientations \cite{knill2013dirac}. \newline

As future outlook, a natural field theoretical extension is the inclusion of self-interactions. One could follow  self-consistent methods inspired from Dynamical Mean Field Theory \cite{georges1996dynamical} to calculate the spectrum and higher $n$-point functions. 
Another question is to evaluate the asymptotics of the off-diagonal generating function of the transition weights, in order to compute the exponential factor of the associated heat kernel. 

There are a few further research avenues that motivated our random processes on trees. 
First, trees stand as the most elementary lowest dimensional structure that can admit some discrete probability measure. 
They also enjoy full solvability from recurrence relations as done with Galton-Watson trees \cite{durhuus2007spectral}.
Secondly, tensor models \cite{guruau2017random}, whose Feynman diagram duals provide a discretization of piecewise-linear manifolds at finite $N$, display in their large-$N$ limit a branching behavior of the type of Galton-Watson trees, with Hausdorff $d_H=2$ and spectral dimension $d_s=4/3$ \cite{Gurau_2013}. 
Can we relate couplings between additional fields and random tensors to a random process on such trees?
Another interesting line of thought is the application of our results to a discrete version of holography, $p$-adic AdS/CFT \cite{Gubser_2017}, where the boundary is the $p$-adic line and the Bethe lattice (also called the Bruhat-Tits tree) stands for the bulk. 
Spin fields in the bulk were first considered in \cite{Gubser_2019}. 
Their Fermionic part of the action is similar to ours, but they include a gauge field as well.
Field theory on the $p$-adic numbers has attracted a lot of interest due to its solvability inherited from its hierarchical structure, see for example \cite{meurice1991symanzik,abdesselam2013rigorous}.
Finally, an obviously important problem is to continue our exploration of the Dirac walk on more general graphs that involve loops (higher complexes, random regular, Erd\"os-Renyi, etc.) to see if this walk is at all related to the standard notion of Fermions. Recalling the generalization of the Dirac operator $\sD$ to higher order complexes, introduced in \cite{Bianconi_2021,Bianconi_2023,Bianconi_2024}, it would be interesting to search for a combinatorial interpretation of such Dirac operators, that could generalize the expansions over spanning forests \cite{Caracciolo:2004hz}.

\section*{Acknowledgements}
The authors would like to thank Jan Ambj\o{}rn, Rudrajit Banerjee, Ginestra Bianconi, Shinobu Hikami, Philipp Hoehn, David Meyer, Yasha Neiman, and Makiko Sasada.
We would also like to thank the Insitute Henri Poincar\'e thematic program ``Quantum Gravity, Random Geometry and Holography", 9 January - 8 February 2023, where part of the work was done and many useful discussions took place.

\let\oldbibliography\thebibliography 
\renewcommand{\thebibliography}[1]{\oldbibliography{#1}
\setlength{\itemsep}{-1pt}}

\begin{thebibliography}{10}

\bibitem{Loll_2019}
R.~Loll, \emph{Quantum gravity from causal dynamical triangulations: a review}, \href{https://doi.org/10.1088/1361-6382/ab57c7}{\emph{Classical and Quantum Gravity} {\bfseries 37} (2019) 013002}.

\bibitem{ambjorn2022elementary}
J.~Ambjorn, \emph{Elementary Introduction to Quantum Geometry}, CRC Press (2022).

\bibitem{deboer2022frontiers}
J.~de~Boer et~al., \emph{{Frontiers of Quantum Gravity: shared challenges, converging directions}},  \href{https://arxiv.org/abs/2207.10618}{{\ttfamily 2207.10618}}.

\bibitem{Williams:2006kp}
R.M.~Williams, \emph{{Discrete quantum gravity}}, \href{https://doi.org/10.1088/1742-6596/33/1/004}{\emph{J. Phys. Conf. Ser.} {\bfseries 33} (2006) 38}.

\bibitem{anderson1958absence}
P.W.~Anderson, \emph{Absence of diffusion in certain random lattices}, \href{https://doi.org/10.1103/PhysRev.109.1492}{\emph{Phys. Rev.} {\bfseries 109} (1958) 1492}.

\bibitem{David:1985nj}
F.~David, \emph{{A Model of Random Surfaces with Nontrivial Critical Behavior}}, \href{https://doi.org/10.1016/0550-3213(85)90363-3}{\emph{Nucl. Phys. B} {\bfseries 257} (1985) 543}.

\bibitem{Knizhnik:1988ak}
V.G.~Knizhnik, A.M.~Polyakov and A.B.~Zamolodchikov, \emph{{Fractal Structure of 2D Quantum Gravity}}, \href{https://doi.org/10.1142/S0217732388000982}{\emph{Mod. Phys. Lett. A} {\bfseries 3} (1988) 819}.

\bibitem{DISTLER1989509}
J.~Distler and H.~Kawai, \emph{Conformal field theory and 2d quantum gravity}, \href{https://doi.org/https://doi.org/10.1016/0550-3213(89)90354-4}{\emph{Nuclear Physics B} {\bfseries 321} (1989) 509}.

\bibitem{duplantier2011liouville}
B.~Duplantier and S.~Sheffield, \emph{Liouville quantum gravity and {KPZ}}, {\emph{Inventiones mathematicae} {\bfseries 185} (2011) 333} [\href{https://arxiv.org/abs/0808.1560}{{\ttfamily 0808.1560}}].

\bibitem{ding2021introduction}
J.~Ding, J.~Dubedat and E.~Gwynne, \emph{Introduction to the liouville quantum gravity metric},  \href{https://arxiv.org/abs/2109.01252}{{\ttfamily 2109.01252}}.

\bibitem{Symanzik:1968zz}
K.~Symanzik, \emph{{Euclidean quantum field theory}}, {\emph{Conf. Proc. C} {\bfseries 680812} (1968) 152}.

\bibitem{jaffe2000constructive}
A.~Jaffe, \emph{Constructive quantum field theory}, \href{https://doi.org/https://doi.org/10.1142/9781848160224_0007}{\emph{Mathematical physics} {\bfseries 5} (2000) 111}.

\bibitem{aizenman2021marginal}
M.~Aizenman and H.~Duminil-Copin, \emph{Marginal triviality of the scaling limits of critical 4d {Ising} and $\lambda\phi^4$ models}, \href{https://doi.org/https://doi.org/10.4007/annals.2021.194.1.3}{\emph{Annals of Mathematics} {\bfseries 194} (2021) 163}.

\bibitem{fernandez2013random}
R.~Fern{\'a}ndez, J.~Fr{\"o}hlich and A.D.~Sokal, \emph{Random walks, critical phenomena, and triviality in quantum field theory}, Springer Science \& Business Media (2013).

\bibitem{lawler12}
G.F.~Lawler, ``Random walk problems motivated by statistical physics.'' \url{http://www.math.uchicago.edu/~lawler/russia.pdf}.

\bibitem{Bodmann_1999}
B.~Bodmann, H.~Leschke and S.~Warzel, \emph{A rigorous path integral for quantum spin using flat-space {Wiener} regularization}, \href{https://doi.org/10.1063/1.532714}{\emph{Journal of Mathematical Physics} {\bfseries 40} (1999) 2549–2559}.

\bibitem{jaroszewicz1994random}
T.~Jaroszewicz and P.S.~Kurzepa, \emph{Random walk representations and four-fermion interactions}, \href{https://doi.org/10.1006/aphy.1994.1017}{\emph{Annals of Physics (New York); (United States)} {\bfseries 230} (1994) }.

\bibitem{alexander1982density}
S.~Alexander and R.~Orbach, \emph{Density of states on fractals:``fractons"}, \href{https://doi.org/doi = {10.1051/jphyslet:019820043017062500}}{\emph{Journal de Physique Lettres} {\bfseries 43} (1982) 625}.

\bibitem{kozma2009alexander}
G.~Kozma and A.~Nachmias, \emph{The {Alexander}-{Orbach} conjecture holds in high dimensions}, \href{https://doi.org/https://doi.org/10.1007/s00222-009-0208-4}{\emph{Inventiones mathematicae} {\bfseries 178} (2009) 635}.

\bibitem{barlow2006random}
M.T.~Barlow and T.~Kumagai, \emph{Random walk on the incipient infinite cluster on trees}, \href{https://doi.org/DOI: 10.1215/ijm/1258059469}{\emph{Illinois Journal of Mathematics} {\bfseries 50} (2006) 33}.

\bibitem{burioni1996universal}
R.~Burioni and D.~Cassi, \emph{Universal properties of spectral dimension}, \href{https://doi.org/10.1103/PhysRevLett.76.1091}{\emph{Phys. Rev. Lett.} {\bfseries 76} (1996) 1091}.

\bibitem{delporte2021perturbative}
N.~Delporte and V.~Rivasseau, \emph{Perturbative quantum field theory on random trees}, \href{https://doi.org/10.1007/s00220-020-03874-2}{\emph{Communications in Mathematical Physics} {\bfseries 381} (2021) 857}.

\bibitem{montvay1994quantum}
I.~Montvay and G.~M{\"u}nster, \emph{Quantum fields on a lattice}, Cambridge University Press (1994).

\bibitem{Caracciolo:2004hz}
S.~Caracciolo, J.L.~Jacobsen, H.~Saleur, A.D.~Sokal and A.~Sportiello, \emph{{Fermionic field theory for trees and forests}}, \href{https://doi.org/10.1103/PhysRevLett.93.080601}{\emph{Phys. Rev. Lett.} {\bfseries 93} (2004) 080601} [\href{https://arxiv.org/abs/cond-mat/0403271}{{\ttfamily cond-mat/0403271}}].

\bibitem{knill2013dirac}
O.~Knill, \emph{The {Dirac} operator of a graph}, \href{https://doi.org/10.48550/arXiv.1306.2166}{\emph{arXiv:1306.2166} }.

\bibitem{catterall2018kahler}
S.~Catterall, J.~Laiho and J.~Unmuth-Yockey, \emph{K\"ahler-{Dirac} fermions on {Euclidean} dynamical triangulations}, \href{https://doi.org/10.1103/PhysRevD.98.114503}{\emph{Physical Review D} {\bfseries 98} (2018) 114503}.

\bibitem{banks1982geometric}
T.~Banks, Y.~Dothan and D.~Horn, \emph{Geometric fermions}, \href{https://doi.org/10.1016/0370-2693(82)90571-8}{\emph{Physics Letters B} {\bfseries 117} (1982) 413}.

\bibitem{casiday2022laplace}
B.~Casiday, I.~Contreras, T.~Meyer, S.~Mi and E.~Spingarn, \emph{Laplace and {Dirac} operators on graphs}, \href{https://doi.org/10.48550/arXiv.2203.02782}{\emph{Linear and Multilinear Algebra} (2022) 1}.

\bibitem{baxter1982exactly}
R.J.~Baxter, \emph{Exactly solved models in statistical mechanics}, Academic Press (1982).

\bibitem{pascazio2023anderson}
S.~Pascazio, A.~Scardicchio and M.~Tarzia, \emph{Anderson localization on the {Bethe} lattice},  in \emph{Spin Glass Theory and Far Beyond: Replica Symmetry Breaking After 40 Years}, pp.~335--352, World Scientific (2023).

\bibitem{rivoire2004glass}
O.~Rivoire, G.~Biroli, O.C.~Martin and M.~M{\'e}zard, \emph{Glass models on {Bethe} lattices}, \href{https://doi.org/10.1140/epjb/e2004-00030-4}{\emph{The European Physical Journal B-Condensed Matter and Complex Systems} {\bfseries 37} (2004) 55}.

\bibitem{mosseri1982bethe}
R.~Mosseri and J.~Sadoc, \emph{The {Bethe} lattice: a regular tiling of the hyperbolic plane}, \href{https://doi.org/10.1051/jphyslet:01982004308024900}{\emph{Journal de Physique Lettres} {\bfseries 43} (1982) 249}.

\bibitem{Breuckmann_2020}
N.P.~Breuckmann, B.~Placke and A.~Roy, \emph{Critical properties of the {Ising} model in hyperbolic space}, \href{https://doi.org/10.1103/physreve.101.022124}{\emph{Physical Review E} {\bfseries 101} (2020) }.

\bibitem{wu2000ising}
C.C.~Wu, \emph{Ising models on hyperbolic graphs {II}}, \href{https://doi.org/10.1023/A:1018763008810}{\emph{Journal of Statistical Physics} {\bfseries 100} (2000) 893}.

\bibitem{gandolfo2015manifold}
D.~Gandolfo, J.~Ruiz and S.~Shlosman, \emph{A manifold of pure {Gibbs} states of the {Ising} model on the {Lobachevsky} plane}, \href{https://doi.org/10.1007/s00220-014-2136-4}{\emph{Communications in Mathematical Physics} {\bfseries 334} (2015) 313}.

\bibitem{hughes1982random}
B.D.~Hughes and M.~Sahimi, \emph{Random walks on the {Bethe} lattice}, \href{https://doi.org/10.1007/BF01011791}{\emph{Journal of Statistical Physics} {\bfseries 29} (1982) 781}.

\bibitem{monthus1996random}
C.~Monthus and C.~Texier, \emph{Random walk on the {Bethe} lattice and hyperbolic {Brownian} motion}, \href{https://doi.org/10.1088/0305-4470/29/10/019}{\emph{Journal of Physics A: Mathematical and General} {\bfseries 29} (1996) 2399}.

\bibitem{Bianconi_2021}
G.~Bianconi, \emph{The topological dirac equation of networks and simplicial complexes}, \href{https://doi.org/10.1088/2632-072X/ac19be}{\emph{Journal of Physics: Complexity} {\bfseries 2} (2021) 035022}.

\bibitem{Bianconi_2023}
G.~Bianconi, \emph{Dirac gauge theory for topological spinors in 3+1 dimensional networks}, \href{https://doi.org/10.1088/1751-8121/acdc6a}{\emph{Journal of Physics A: Mathematical and Theoretical} {\bfseries 56} (2023) 275001}.

\bibitem{Bianconi_2024}
G.~Bianconi, \emph{The mass of simple and higher-order networks}, \href{https://doi.org/10.1088/1751-8121/ad0fb5}{\emph{Journal of Physics A: Mathematical and Theoretical} {\bfseries 57} (2023) 015001}.

\bibitem{nokkala2023complex}
J.~Nokkala, J.~Piilo and G.~Bianconi, \emph{Complex quantum networks: a topical review},  2023.
\newblock 10.48550/arXiv.2311.16265.

\bibitem{Becher:1982ud}
P.~Becher and H.~Joos, \emph{{The Dirac-Kahler Equation and Fermions on the Lattice}}, \href{https://doi.org/10.1007/BF01614426}{\emph{Z. Phys. C} {\bfseries 15} (1982) 343}.

\bibitem{lim2019hodge}
L.-H.~Lim, \emph{Hodge laplacians on graphs},  2019.

\bibitem{norris1998markov}
J.~Norris, \emph{Markov Chains}, Cambridge Series in Statistical and Probabilistic Mathematics, Cambridge University Press (1998).

\bibitem{gurau2014renormalization}
R.~Gur{\u{a}}u, V.~Rivasseau and A.~Sfondrini, \emph{{Renormalization: an advanced overview}},  \href{https://arxiv.org/abs/1401.5003}{{\ttfamily 1401.5003}}.

\bibitem{friedrich2000dirac}
T.~Friedrich, \emph{Dirac operators in {Riemannian} geometry}, vol.~25, American Mathematical Soc. (2000).

\bibitem{Montvay:1994cy}
I.~Montvay and G.~Munster, \emph{{Quantum fields on a lattice}}, Cambridge Monographs on Mathematical Physics, Cambridge University Press (3, 1997), \href{https://doi.org/10.1017/CBO9780511470783}{10.1017/CBO9780511470783}.

\bibitem{Kahler1962}
E.~K\"ahler, \emph{Der innere differentialkalk\"ul}, \href{https://doi.org/10.1007/BF02992927}{\emph{Abhandlungen aus dem Mathematischen Seminar der Universit\"at Hamburg} {\bfseries 25} (1962) 192}.

\bibitem{Cassidy:https://doi.org/10.48550/arxiv.2203.02782}
B.~Casiday, I.~Contreras, T.~Meyer, S.~Mi and E.~Spingarn, \emph{Laplace and dirac operators on graphs}, {\emph{Linear and Multilinear Algebra} (2022) 1} [\href{https://arxiv.org/abs/2203.02782}{{\ttfamily 2203.02782}}].

\bibitem{Knill:https://doi.org/10.48550/arxiv.1306.2166}
O.~Knill, \emph{The dirac operator of a graph},  \href{https://arxiv.org/abs/1306.2166}{{\ttfamily 1306.2166}}.

\bibitem{Peskin:1995ev}
M.E.~Peskin and D.V.~Schroeder, \emph{{An Introduction to quantum field theory}}, Addison-Wesley, Reading, USA (1995).

\bibitem{Monthus_1996}
C.~Monthus and C.~Texier, \emph{Random walk on the {Bethe} lattice and hyperbolic brownian motion}, \href{https://doi.org/10.1088/0305-4470/29/10/019}{\emph{Journal of Physics A: Mathematical and General} {\bfseries 29} (1996) 2399}.

\bibitem{Hughes1982}
B.D.~Hughes and M.~Sahimi, \emph{Random walks on the {Bethe} lattice}, \href{https://doi.org/10.1007/BF01011791}{\emph{Journal of Statistical Physics} {\bfseries 29} (1982) 781}.

\bibitem{Correia_1998}
J.D.~Correia and J.F.~Wheater, \emph{The spectral dimension of non-generic branched polymer ensembles}, \href{https://doi.org/10.1016/s0370-2693(98)00055-0}{\emph{Physics Letters B} {\bfseries 422} (1998) 76–81}.

\bibitem{Ambj_rn_2005}
J.~Ambjørn, J.~Jurkiewicz and R.~Loll, \emph{The spectral dimension of the universe is scale dependent}, \href{https://doi.org/10.1103/physrevlett.95.171301}{\emph{Physical Review Letters} {\bfseries 95} (2005) }.

\bibitem{10.5555/1506267}
P.~Flajolet and R.~Sedgewick, \emph{Analytic Combinatorics}, Cambridge University Press, USA, 1~ed. (2009).

\bibitem{economou2006green}
E.~Economou, \emph{Green's Functions in Quantum Physics}, Springer Series in Solid-State Sciences, Springer (2006).

\bibitem{georges1996dynamical}
A.~Georges, G.~Kotliar, W.~Krauth and M.J.~Rozenberg, \emph{{Dynamical mean-field theory of strongly correlated fermion systems and the limit of infinite dimensions}}, \href{https://doi.org/10.1103/RevModPhys.68.13}{\emph{Rev. Mod. Phys.} {\bfseries 68} (1996) 13}.

\bibitem{durhuus2007spectral}
B.~Durhuus, T.~Jonsson and J.F.~Wheater, \emph{{The Spectral dimension of generic trees}}, \href{https://doi.org/10.1007/s10955-007-9348-3}{\emph{J. Statist. Phys.} {\bfseries 128} (2007) 1237} [\href{https://arxiv.org/abs/math-ph/0607020}{{\ttfamily math-ph/0607020}}].

\bibitem{guruau2017random}
R.G.~Gur{\u{a}}u, \emph{Random tensors}, Oxford University Press (2017).

\bibitem{Gurau_2013}
R.~Gurau and J.P.~Ryan, \emph{Melons are branched polymers}, \href{https://doi.org/10.1007/s00023-013-0291-3}{\emph{Annales Henri Poincar{\'{e}}} {\bfseries 15} (2013) 2085}.

\bibitem{Gubser_2017}
S.S.~Gubser, J.~Knaute, S.~Parikh, A.~Samberg and P.~Witaszczyk, \emph{p-adic ads/cft}, \href{https://doi.org/10.1007/s00220-016-2813-6}{\emph{Communications in Mathematical Physics} {\bfseries 352} (2017) 1019–1059}.

\bibitem{Gubser_2019}
S.S.~Gubser, C.~Jepsen and B.~Trundy, \emph{Spin in $p$-adic {AdS}/{CFT}}, \href{https://doi.org/10.1088/1751-8121/ab0757}{\emph{Journal of Physics A: Mathematical and Theoretical} {\bfseries 52} (2019) 144004}.

\bibitem{meurice1991symanzik}
Y.~Meurice, \emph{{A path integral formulation of p-adic quantum mechanics}}, \href{https://doi.org/10.1016/0370-2693(90)90171-2}{\emph{Phys. Lett. B} {\bfseries 245} (1990) 99}.

\bibitem{abdesselam2013rigorous}
A.~Abdesselam, A.~Chandra and G.~Guadagni, \emph{Rigorous quantum field theory functional integrals over the p-adics {I}: anomalous dimensions},  \href{https://arxiv.org/abs/[1302.5971]}{{\ttfamily [1302.5971]}}.

\end{thebibliography}

\providecommand{\href}[2]{#2}\begingroup\raggedright\endgroup
\end{document}